\documentclass[prb,twocolumn,groupedaddress,showpacs]{revtex4}
	
\usepackage{graphicx,amsmath,appendix,simplewick}
\newcommand{\sgn}{\mathrm{sgn}}

\begin{document}

\title{More Realistic Hamiltonians for the Fractional Quantum Hall Regime
in GaAs and Graphene}

\author{Michael R. Peterson$^{1,2}$ and Chetan Nayak$^{2,3}$}
\affiliation{$^{1}$Department of Physics \& Astronomy, California State University Long Beach,  Long Beach, California 90840, USA}
\affiliation{$^{2}$Department of Physics, University of California, Santa Barbara, California 93106, USA}
\affiliation{$^{3}$Microsoft Research, Station Q, Elings Hall, University of California, Santa Barbara, California 93106, USA}

\begin{abstract}
We construct an effective Hamiltonian for electrons in the fractional quantum 
Hall regime for GaAs and graphene that takes into account 
Landau level mixing (for both GaAs and graphene) and sub-band mixing (for GaAs,
due to the non-zero-width of the quantum well).
This mixing has the important qualitative effect
of breaking particle-hole symmetry as well as renormalizing the 
strength of the inter-particle interactions. Both effects could have
important consequences for the prospect that
the fractional quantum Hall effect at $\nu=5/2$ is described by
states that support non-Abelian excitations such as the
Moore-Read Pfaffian or anti-Pfaffian states.  
For GaAs, Landau level and sub-band mixing break particle-hole 
symmetry in \textit{all} Landau levels and sub-band mixing, due to finite-thickness, causes additional short-distance softening of the Coulomb interaction, further renormalizing
the Hamiltonian--additionally, the Landau level and sub-band energy spacings are comparable so it is crucial to consider both 
effects simultaneously.  We find that in graphene, Landau-level mixing only
breaks particle-hole symmetry outside of the lowest
Landau level ($N\neq 0$).  Landau level mixing is likely to be especially important 
in graphene since the Landau-level mixing parameter is independent of the
external magnetic field and is of order one.  Our realistic Hamiltonians will serve as starting points for future numerical studies.
\end{abstract}

\date{\today}


\maketitle

\section{INTRODUCTION and Motivation}
\label{intro}

One of the outstanding experimental challenges
in physics  is to determine whether the
fractional quantum Hall effect
supports
non-Abelian anyon excitations. An affirmative answer
would constitute the discovery of a new type of, so far unobserved, particle and
could pave the way for topologically-protected
quantum information processing \cite{Kitaev97,DasSarma05,Nayak08}.

The most promising and notable system for which the search for non-Abelian anyons is 
taking place is in the $\nu=5/2$
fractional quantum Hall effect~\cite{Willett87,Pan99b,Eisenstein02,Radu08,Dolev08,Willett09,Stern10,Bid10,Rhone10,Venkatachalam11,Tiemann11}.
The search is motivated by two conjectures. The first is that the
Moore-Read Pfaffian state \cite{Moore91} and the anti-Pfaffian
state \cite{Lee07,Levin07} are representatives of universality classes
that have non-Abelian anyon excitations. This conjecture
has been recently shown to be true \cite{Bonderson11a}
(see also Refs. \onlinecite{Nayak96c,Read96,Gurarie97,Read00,Ivanov01,Tserkovnyak03,Stern04,Stone06,Seidel08,Read08,Baraban09,Prodan09}).
The second conjecture is that the
experimentally observed state of matter responsible for the $\nu=5/2$ 
fractional quantum Hall effect (FQHE)
is in one of these two universality classes.  Several
numerical studies \cite{Morf98,Rezayi00,Feiguin08,Peterson08,Peterson08b,Peterson12,Rezayi09,Wojs10,Storni10}
have provided evidence supporting this conjecture by showing that
the ground states of simplified model Hamiltonians 
are in these two universality classes (Moore-Read Pfaffian or anti-Pfaffian).
Even though these model Hamiltonians  are physically reasonable, a number of studies \cite{Rezayi00,Peterson08,Peterson08b,Wojs10}
highlight the sensitivity of numerical results\cite{Morf98,Rezayi00,Feiguin08,Peterson08,Peterson08b,Rezayi09,Wojs10,Storni10}
to the parameters of these `toy model'
Hamiltonians (and to some degree, the system size).
Experiments have found that relatively minor modifications to 
the system such as transposition of the half-filled Landau level from $\nu=5/2$
to $\nu=9/2, 11/2, \ldots$ or to $\nu=1/2, 3/2$ lead to metallic states that are
anisotropic  \cite{Lilly99a,Du99} or isotropic and Fermi liquid-like \cite{Willett93}. Similar anisotropic or isotropic metallic phases are also observed
at $\nu=5/2$  upon application of, respectively, a small\cite{Pan99a,Lilly99b} or large \cite{Xia11} in-plane magnetic field (compared 
in magnitude to the perpendicular magnetic field).
Thus, there are at least four different phases (and, perhaps, many more)
which can occur at $\nu=5/2$, depending on the details of the Hamiltonian.
Thus, we are encouraged to ask basic questions which remain unanswered: 
is the ground state of a more {\it realistic} model Hamiltonian for
$\nu=5/2$ in one of these two
non-Abelian universality classes? What is the quantum phase diagram of a realistic Hamiltonian for $\nu=5/2$?

The sensitivity of the $\nu=5/2$ system to small changes in the Hamiltonian
is a cause for concern since the canonical simplified model ignores
both Landau-level mixing and the non-zero-width of the quasi-two-dimensional (quasi-2D) electron system.
(The exceptions are Refs.~\onlinecite{Rezayi09,Wojs10}, which include
Landau-level mixing and Refs.~\onlinecite{Peterson08,Peterson08b,Peterson12},
which include non-zero-width.) Landau level mixing and non-zero-width
can be neglected if the Landau level mixing parameter
$\kappa=({e^2}/\epsilon\ell_0)/\hbar\omega_c \ll 1$ and
the non-zero-width of the quantum-well $d/\ell_0 \ll 1$, respectively.  Here,
$e^2/\epsilon\ell_0$ is the scale of the Coulomb energy, 
$\omega_c$ is the cyclotron frequency,
and $\ell_0=\sqrt{\hbar c/eB}$ is the magnetic length,
so the dimensionless parameter $\kappa$ is
given by $\kappa\approx 2.52/\sqrt{B[\mathrm{T}]}$.
Therefore, for experiments at magnetic fields in the
range $1-10$T, $\kappa$ is in the range $0.8-2.5$.  Similarly, for most GaAs samples $d/\ell_0\approx 2-3$.
It is not obvious that either one of these parameters can be considered small and, in fact, for common 
experimental parameters the Landau-level  and sub-band energy spacing are comparable.  Thus, it is potentially dangerous 
to consider one effect and not the other since, \textit{a priori}, both effects are of approximately equal importance.  Given that the ground state at $\nu=5/2$ depends sensitively
on the precise Hamiltonian and that the Hamiltonians studied in
all previous numerical results neglect potentially important effects, one may question 
their  connection to experiments~\cite{footnote1,footnote2}.  

In fact, recently the FQHE at $\nu=5/2$ 
has been studied experimentally in the regime of strong Landau level mixing ($\kappa>2.5$) and interesting 
non-linear behavior of the FQHE energy gap has been observed as a function of $\kappa$ (or density)~\cite{Samkharadze13}.  
Thus, in order to make meaningful experimental predictions for the FQHE in GaAs (or generally any two-dimensional electron gas),
it is necessary to study realistic Hamiltonians
which include the effects of both Landau level mixing and non-zero-width.
This is true not only at $\nu=5/2$, but throughout
the $N=1$ Landau level, where Landau level mixing is a factor.

The scenario described above for GaAs is potentially an even more pressing issue when
one considers the FQHE in graphene.  Due to the 
linear dispersion of electrons in graphene,
the cyclotron energy is $\sgn(N)\sqrt{2|N|}\hbar v_F/\ell_0$, compared 
to $\hbar\omega(N+1/2)$ for GaAs.  (In GaAs, $N=0$, 1, 2,$\ldots$ whereas 
in graphene $-\infty<N<\infty$).
This means that the Landau level mixing parameter in graphene 
$\tilde\kappa=(e^2/\epsilon\ell_0)/(\hbar v_F/\ell_0)=
e^2/(\epsilon \hbar v_F)$ is,
interestingly, \textit{independent} of magnetic field strength.
In addition, the spacing between Landau levels 
varies as $~1/\sqrt{2N}$ for large $N$--the 
Landau levels get closer together in energy the higher up, or lower down
below $N=0$, one goes.  Consequently, the effective Landau level mixing parameter for
the $N^{\rm th}$ Landau level is $\propto\tilde\kappa\sqrt{N}$, which
increases with increasing Landau level. 

Experimentally, the FQHE has been observed in graphene in suspended samples~\cite{Bolotin09,Ghahari11,Du09} (i.e., free standing graphene)
and on a boron nitride substrate~\cite{Dean11}. In suspended graphene, the Landau level mixing parameter is approximately $\tilde\kappa\approx 2.2$ 
whereas on a boron nitride substrate it is much lower
due to the reduction in dielectric constant: $\tilde\kappa\approx 0.5-0.8$.
In both cases $\tilde\kappa$ is obviously not small so
it would appear that there is no experimental observation
of the FQHE in graphene for which one can ignore Landau level mixing.
Many theoretical works have considered the 
FQHE in graphene~\cite{Apalkov06,Toke06,Toke07a,Goerbig07,Papic09,Toke07b},
pointing  out the similarities and differences between the FQHE in GaAs and
graphene, but none  have taken into account the effect of Landau level mixing.

Therefore, in this paper we construct and study a realistic Hamiltonian for GaAs 
in the lowest two Landau levels ($N=0$ and $N=1$; no FQHE has been observed in $N\geq 2$) that takes into account 
Landau level and sub-band mixing. 
In addition, we construct a realistic Hamiltonian for graphene in the $N=0,\pm1,\pm2$ Landau levels which includes Landau level mixing effects which have, so 
far, been totally ignored.  
Although we are emphasizing the experimental systems of GaAs and graphene in our calculations, they are appropriate for \textit{any} fermionic system
with either parabolic (GaAs) or linear (graphene) bands
up to the specific experimental parameters.
Once these effective Hamiltonians are characterized, they can be used to study a variety of problems in which sub-band and/or Landau 
level mixing effects may play a prominent role.

Our starting point is the systematic treatment of
Landau-level mixing formulated in Ref.~\onlinecite{Bishara09a}.
The effective Hamiltonian for electrons in the $N^{\rm th}$ Landau level
can be derived by integrating out all other Landau levels via an expansion in powers
of $\kappa$(or $\tilde\kappa$). This was done~\cite{Bishara09a} for GaAs for zero width,
but the same procedure allows us to integrate out higher sub-bands in the
same way as higher Landau levels and/or consider graphene, as we discuss below.

Crucially, we find an error in the normal-ordering of the effective
Hamiltonian of the previous work~\cite{Bishara09a}. Correcting this error,
we find that the renormalization of the two-body interaction is
significantly modified and we examine the implications.

We emphasize that, for GaAs, we include the non-zero width
of the 2D layer, as a result of which the effective interaction between electrons
is ``softened" at short distances since the single-particle wave functions are smeared out
over a length scale $d$ in the direction perpendicular to the plane.
This effect was found to stabilize the Moore-Read Pfaffian and anti-Pfaffian
states\cite{Peterson08,Peterson08b} in the $N=1$ Landau level, unlike
in the lowest Landau level, where it weakens the FQHE.
A further effect of non-zero-width
is that electron-electron interactions can cause mixing with the
higher quantum well sub-bands corresponding to motion in the direction perpendicular to the 
two-dimensional plane, in a manner analogous to Landau-level mixing.
(Strictly speaking, a system with only sub-band mixing 
could also cause a breaking of particle-hole symmetry.
However, Landau-level mixing is generally a stronger effect than sub-band mixing.)
We take these two facets of non-zero width (the softening of the
``bare" Coulomb interaction and mixing with higher sub-bands) into account
in a systematic manner. This has not, to the best of our knowledge,
been previously done.  Of course, for graphene, the effective width
of the 2D graphene sheet is negligible, so there are no sub-bands 
that need to be considered.

Strictly speaking, our Hamiltonian can only be justifiably
called ``realistic" for small values of $\kappa$ (or $\tilde\kappa$).
For GaAs, small $\kappa$ corresponds to 
higher values of the magnetic field whereas for graphene,
small $\tilde\kappa$ corresponds to a substrate with high dielectric constant. 
However, we find that the coefficients in the expansion in powers of $\kappa$
are small in GaAs (even with the correction mentioned
in the previous paragraph) and, consequently,
our Hamiltonian may even be realistic when $\kappa$  is not small.  For graphene, the 
coefficients in the $N=0$ Landau level are small, similar to those in GaAs,
but for $N\neq 0$, this is no longer the case.
Thus, these expansions may be valid even in the regime
where most experimental observations of the FQHE at $\nu=5/2$ in
GaAs have taken place ($0.7 < \kappa < 1.8$),   
as well as where the FQHE has been experimentally 
observed in graphene~\cite{Bolotin09,Ghahari11,Du09,Dean11,DasSarma-RMP2011} ($0.5<\tilde\kappa<2.2$).   
Our calculations might also help explain some of the peculiarities of the graphene FQHE observations,
such as the fact that the FQHE has been observed 
in only the lowest Landau level (see Refs.~\onlinecite{Bolotin09,Ghahari11,Du09,Dean11}).  
This point will be discussed further below.

\textit{Previous results for GaAs}:
The Landau-level mixing Hamiltonian of Ref. \onlinecite{Bishara09a}
for GaAs was studied for zero width at $\nu=5/2$
by Wojs {\it et al.} \cite{Wojs10} using exact diagonalization in the spherical geometry,
and it was concluded that, over nearly the entire range of Landau level mixing
of experimental interest, $0<\kappa\leq 3$,
the overlap between the ground state and the Moore-Read Pfaffian state
was larger than either the overlap between the ground state 
and the composite fermion Fermi-liquid-like wavefunction\cite{footnote3} or the
overlap between the ground state  and the
anti-Pfaffian wavefunction.
However, the computation of Wojs {\it et al.}\cite{Wojs10} 
did not take finite-thickness of the 2D system into account;
it is not clear that these overlaps at different values of the magnetic flux track the (extrapolated)
ground state energies, which is what determines the true ground state;
and, more importantly, the Hamiltonian did not contain the corrected
normal-ordering of the three-body term.  
Rezayi and Simon\cite{Rezayi09} did a similar study using the torus geometry 
but they did not use the Hamiltonian of Ref. \onlinecite{Bishara09a}
and instead simulated Landau level mixing by diagonalizing 
in a truncated Hilbert space.
In contrast to Wojs {\it et al.}\cite{Wojs10}, 
they found the anti-Pfaffian to have a higher overlap with
the exact ground state.  Even though both studies 
used wavefunction overlap to measure to which universality
class the ground state belonged, it is difficult 
to directly compare their contrasting results since they used
different Hamiltonians and different geometries.
Of course, it is possible that the previous error in 
the normal ordering of the three-body term is the origin of these
apparently contradictory results.

We point out that there has been additional previous work that considered both sub-band and Landau level mixing~\cite{PhysRevB.81.035316,Rezayi-PRB1990,PhysRevB.68.113309}.  While many insights can be gained from this work, it is not particularly relevant to this
 study.  For the FQHE, in particular, the effects of both sub-band and Landau level mixing have been studied, either  numerically in the perturbative limit or using a combination of perturbation theory (random 
 phase approximation) and  phenomenological models, however, our calculation 
provides the crucial qualitatively important idea of the breaking of particle-hole symmetry through the generation of three-body terms which were not previously taken into account.

\textit{Previous results for graphene}:
None of the theoretical studies for the FQHE in graphene
mentioned above included the effects of Landau level mixing.

The plan of this paper is as follows.  In Section~\ref{prelims} we review the effective
action description given previously in Ref.~\onlinecite{Bishara09a}.   
To this we add the additional complication of a non-zero-width quantum well.
This is not as simple as just augmenting the Coulomb interaction
because it becomes necessary to integrate out all higher quantum well sub-bands.
We then derive the effective Hamiltonian which follows from this effective
action, paying particular attention to the operator-ordering of the Hamiltonian.
In Sec.~\ref{pseudoVms}, we discuss the effective Hamiltonian which 
includes Landau-level and sub-band mixing to first order in the
Landau level mixing parameter $\kappa$.  In Sec.~\ref{sec-graphene}, we 
consider the effective Hamiltonian for graphene and in
Sec.~\ref{grapheneVms}, we discuss the pseudopotentials.
Finally, in Sec.~\ref{conclusion}, we present 
our conclusions. 

\section{Effective Hamiltonian including landau level
and SUB-BAND MIXING applicable to $\mbox{GaAs}$}
\label{prelims}

\subsection{Effective action}
\label{subsec:effective-action-gaas}

We first begin by discussing the calculation for a system appropriate for GaAs, just to introduce 
the formalism, and then in Sec.~\ref{sec-graphene} we discuss the calculation for a system appropriate to graphene.
Following Ref.~\onlinecite{Bishara09a} we begin with the action for
electrons in a magnetic field:
\begin{multline}
\label{eqn:bare-Hamiltonian}
S = \int \frac{d\omega}{2\pi}\sum_{snm}\bar{c}_{snm \alpha}(\omega)(i\omega-E_{sn}+\mu)c_{snm \alpha}(\omega)\\
-\,\frac{1}{2}\int\prod_{i=1}^4\frac{d\omega_i}{2\pi}V_{43;21}\bar{c}_{s_4 n_4 m_4 \alpha}(\omega_4)\bar{c}_{s_3 n_3 m_3 \beta}(\omega_3)\\
\times c_{s_2 n_2 m_2 \beta}(\omega_2)c_{s_1 n_1 m_1 \alpha}(\omega_1)\\
\times 2\pi \delta(\omega_4+\omega_3-\omega_2-\omega_1)\;.
\end{multline}
Here, $c_{snm \alpha}$ and $\bar{c}_{snm \alpha}$ are Grassmann variables
where $\alpha=\uparrow,\downarrow$ are the 
spin indices; $s=0,1,\ldots$ is the sub-band index; $n=0,1,\ldots$ is the Landau level
index; and $m=0,1,\ldots$ labels orbital states within a Landau level
($m-n$ is the angular momentum).
$E_{sn}=\hbar\omega_c(n+1/2)+\varepsilon_s$ is the cyclotron energy
($\hbar \omega_c = \hbar e B/mc$) plus the sub-band energy $\varepsilon_s$,
and $\mu$ is the chemical potential.   Note that the Zeeman energy has been set to zero
in the above action since it is much smaller than the other energy scales
in the problem.   $V_{43;21}\equiv \langle s_4 n_4 m_4, s_3 n_3 m_3|\hat{V}|s_2 n_2 m_2,s_1 n_1 m_1\rangle$  
is the interaction matrix element, where $\hat{V}$ is 
the electron-electron interaction operator for the
bare two-body Coulomb interaction for electrons in an infinite quantum well of width, $d$, i.e., 
the usual non-zero-width augmented interaction investigated previously~\cite{MacDonald84,He90,Zhang86,Ortalano97,Park98,Park99,Peterson08,Peterson08b,Peterson12}.
Following the standard procedure~\cite{macdonald,jain2007composite}, we define $G_{ab}(k)$ as
\begin{eqnarray}
\label{gab}
G_{ab}(k)=\left(-i\frac{\bar{k}}{\sqrt{2}}\right)^{a-b}\sqrt{\frac{b!}{a!}}L_b^{a-b}(k^2/2)
\end{eqnarray}
for integers $a\geq b$ (note that if $a<b$ we switch $k=k_x+ik_y$ with $\bar{k}=k_x-ik_y$) with $L_a^b(x)$ a generalized Laguerre polynomial.  
We find that the interaction matrix element is
\begin{eqnarray}
\label{eqn:V4321-def}
\label{v}
V_{43,21}&=&\int\frac{d^2k}{2\pi}V_{s_4s_3s_2s_1}(k)e^{-k^2}G_{n_4n_2}(-\bar{k})
G_{n_3n_1}(\bar{k})\nonumber\\&&\times G_{m_4m_2}(-k)G_{m_3m_1}(k)\;,
\end{eqnarray}
where the $4$ in $V_{43,21}$, for example, is short for all the internal degrees of freedom of the fourth particle ($s_4$, $n_4$, and $m_4$) and 
\begin{eqnarray}
\label{vft}
V_{s_4 s_3s_2s_1}(k)&=&\frac{e^2}{2\pi\epsilon l_0}\frac{1}{k}\int dz_1dz_2\phi_{s_4}(z_1)
\phi_{s_3}(z_1)\nonumber\\&&\times\phi_{s_2}(z_2)\phi_{s_1}(z_2)e^{-k|z_1-z_2|}\;,
\end{eqnarray}
where $\phi_{s}(z)$ is the wavefunction for the $z$-dependence of
single-particle wavefunctions in the $s^{\rm th}$ sub-band.
Throughout this work on GaAs we will consider the electrons 
to be confined to an infinite square well of width $d$.
Thus, $\phi_s(z)=\sqrt{2/d}\sin((s+1)\pi z/d)$ 
with $z\in[0,d]$ and $\varepsilon_s (d) = (s+1)^2\pi^2\hbar^2/(2m_z d^2)$ where 
$m_z$ is the effective electron mass in the quantum well.   
The total single particle energy is $E_{sn}=\hbar\omega_c (n+1/2) + \varepsilon_s(d)$. 
If we assume that $m_z$ equals the electron band mass $m$ and write
$d$ in units of the magnetic length $\ell_0=\sqrt{\hbar c/eB}\approx 25\mbox{ nm}/\sqrt{B[\mbox{T}]}$, then
$E_{sn}=\hbar\omega_c[(n+1/2) + (\pi^2/2d^2)(s+1)^2]$.  For 
typical quantum well widths and magnetic fields ($d/\ell_0 \approx 2-4$) the Landau-level spacing and sub-band spacing are 
comparable.  We further note that the energy spacing between sub-bands gets smaller as the inverse square of the quantum well width unlike the 
Landau level spacing which is constant.
Of course, the  potential corresponding to a relevant  experimental system is not an infinite square well~\cite{Liu-PRL2011,Nuebler-PRL2012},
however, the infinite square well captures the key feature of the real
potential~\cite{MacDonald84,He90,Zhang86,Ortalano97,Park98,Park99,Peterson08,Peterson08b,Peterson12}: the ability of electrons to avoid each other by moving in
the $z$-direction, as reflected both by the softening of the
short-distance part of the Coulomb interaction and 
by the possibility of virtual transitions to higher sub-bands (which have nodes
at certain values of $z$).  We caution the teader that there is one qualitative effect not properly captured by modeling the quantum well
with an infinite square well potential; at large enough widths and/or densities, the single layer system can effectively turn into a bilayer.  For the purpose of this work, 
however, we are not considering this situation.

Although the modeling of a more realistic potential is possible within our framework, it is highly dependent on a particular experimental sample.  Hence, 
each sample would require new calculations, which as shown in the following are very laborious.  The main purpose of this work, in this regard, is to make general statements 
and conclusions about the effects of both sub-band and Landau level mixing in a model system.  We reiterate, however, that our formalism  
allows the consideration of  more realistic models of  quantum wells, through the use of local-density-approximation, for example, in a straight-forward manner.  These 
more specific calculations, however, will have to await further study.

We first assume the $N^{\rm th}$ Landau level to be 
partially occupied by electrons and integrate out all
higher/lower Landau levels and sub-bands perturbatively in the 
Coulomb interaction. We thereby generate an effective Hamiltonian
for the $N^{\rm th}$ Landau level which incorporates the effects of virtual
transitions to the other Landau levels and sub-bands.
This effective Hamiltonian is computed perturbatively 
in powers of $\kappa$, where $\kappa=(e^2/\epsilon l_0)/(\hbar\omega_c)$ is the 
Landau level mixing parameter discussed above
($\kappa\approx 2.52\times1 \mbox{T}/\sqrt{B[\mbox{T}]}$).  Thus, 
 $\kappa$ can be varied from very small to very large by changing the magnetic field
 (and the density, in order to keep the filling fraction fixed). This is 
is an important distinction between GaAs and graphene, as we discuss in more
detail below.

At tree level, the effective action is simply
Eq. ~(\ref{eqn:bare-Hamiltonian}) with the sums over Landau levels
restricted to ${n_i}=N$ and the sums over sub-bands restricted to ${s_i}=0$.
Most numerical studies of the quantum Hall effect use the effective
Hamiltonian associated with this effective action,
in which the other Landau levels are integrated out to zeroth order in
the Coulomb interaction or, equivalently, to zeroth order in $\kappa$.
However, if we integrate out the higher Landau levels to
second-order, which amounts
to computing the effective action to first-order in $\kappa$, we obtain:
\begin{eqnarray}
S&=& \int \frac{d\omega}{2\pi}\sum_{m}\bar{c}_{0Nm \alpha}(\omega)(i\omega-E_{0N}+\mu)c_{0Nm \alpha}(\omega)\nonumber\\
&-&\frac{1}{2}\int\prod_{i=1}^4\frac{d\omega_i}{2\pi}u^{(2)}_{43;21}\bar{c}_{0 N m_4 \alpha}(\omega_4)\bar{c}_{0 N m_3 \beta}(\omega_3)\nonumber\\
&&\times c_{0N m_2 \beta}(\omega_2)c_{0N m_1 \alpha}(\omega_1) 2\pi \delta(\omega_4+\omega_3-\omega_2-\omega_1)\nonumber\\
&-&\frac{1}{3!}\int\prod_{i=1}^6\frac{d\omega_i}{2\pi}u^{(3)}_{654;321}\bar{c}_{0 N m_6 \alpha}(\omega_6)\bar{c}_{0 N m_5 \beta}(\omega_5)
\nonumber\\
&&\times\bar{c}_{0 N m_4 \gamma}(\omega_4)
 c_{0N m_3 \gamma}(\omega_3)c_{0N m_2 \beta}(\omega_2)c_{0N m_1 \alpha}(\omega_1)\nonumber\\
&&\times 2\pi \delta(\omega_6+\omega_5+\omega_4-\omega_3-\omega_2-\omega_1)\;.
\end{eqnarray}
As a result of the higher Landau levels which are integrated out,
the bare two-body interaction is renormalized
\begin{eqnarray}
u^{(2)}_{43;21} = V_{43;21} + \kappa \delta u^{(2)}_{43;21}\;
\end{eqnarray}
and a three-body interaction $u^{(3)}_{654;321}$ is generated.
At higher-orders in $\kappa$, four-body, five-body, etc., interactions are
generated, but we will restrict ourselves here to lowest-order in $\kappa$,
so we only need to consider two- and three--body interactions.

\begin{figure}[htbp]
\begin{center}
\includegraphics[width=6.5cm,angle=0]{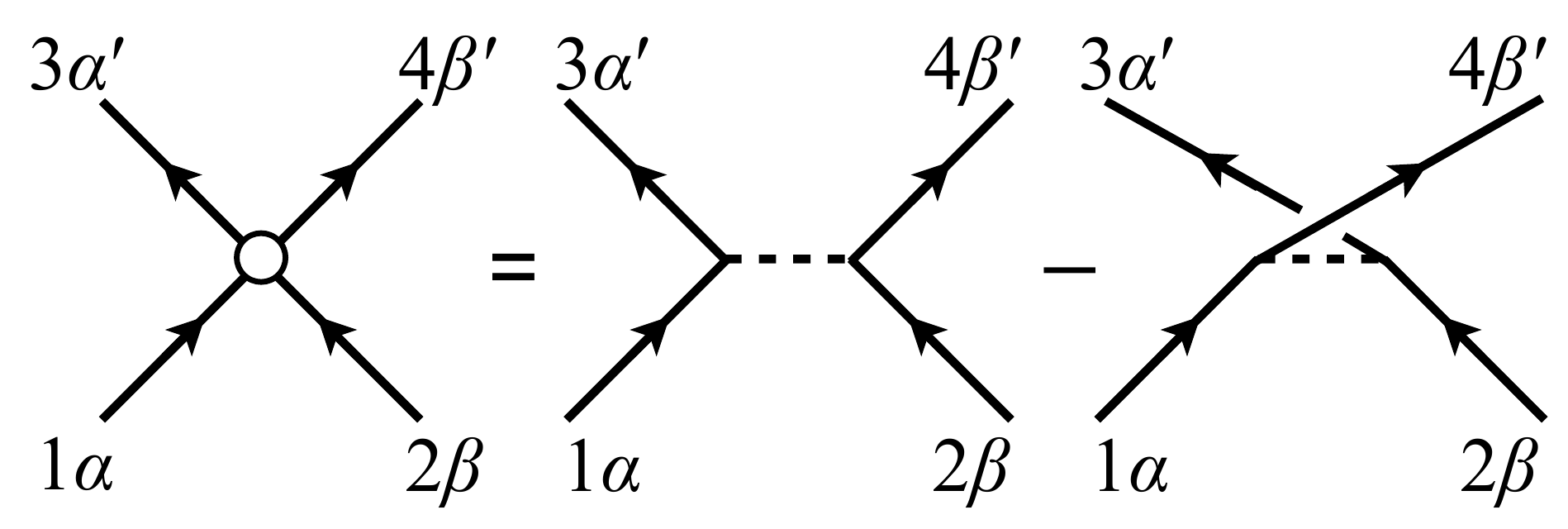}	
\caption{The bare interaction Feynman diagram.  The Coulomb interaction (augmented or not) cannot switch the 
spin so the first term on the right hand side has $\delta^{\alpha\alpha'}\delta^{\beta\beta}$ and the second 
term has $\delta^{\alpha\beta'}\delta^{\beta\alpha'}$.  The labels 1, 2, 3, and 4 correspond to 
$1=\{s_1, n_1, m_1\}$, etc..}
\label{u0}
\end{center}
\end{figure}

The calculation will be presented 
using Feynman diagrams and the single Coulomb vertex
in the diagrammatics is $V^{\beta'\alpha',\beta\alpha}_{43,21}=V_{43,21}\delta^{\alpha\alpha'}\delta^{\beta\beta'} - V_{34,21}\delta^{\alpha\beta'}\delta^{\beta\alpha'}$ where $\alpha$, $\beta$, $\alpha'$, and $\beta'$ label spin indices (see Fig.~\ref{u0}).

\begin{figure}[htbp]
\begin{center}
\includegraphics[width=7.cm,angle=0]{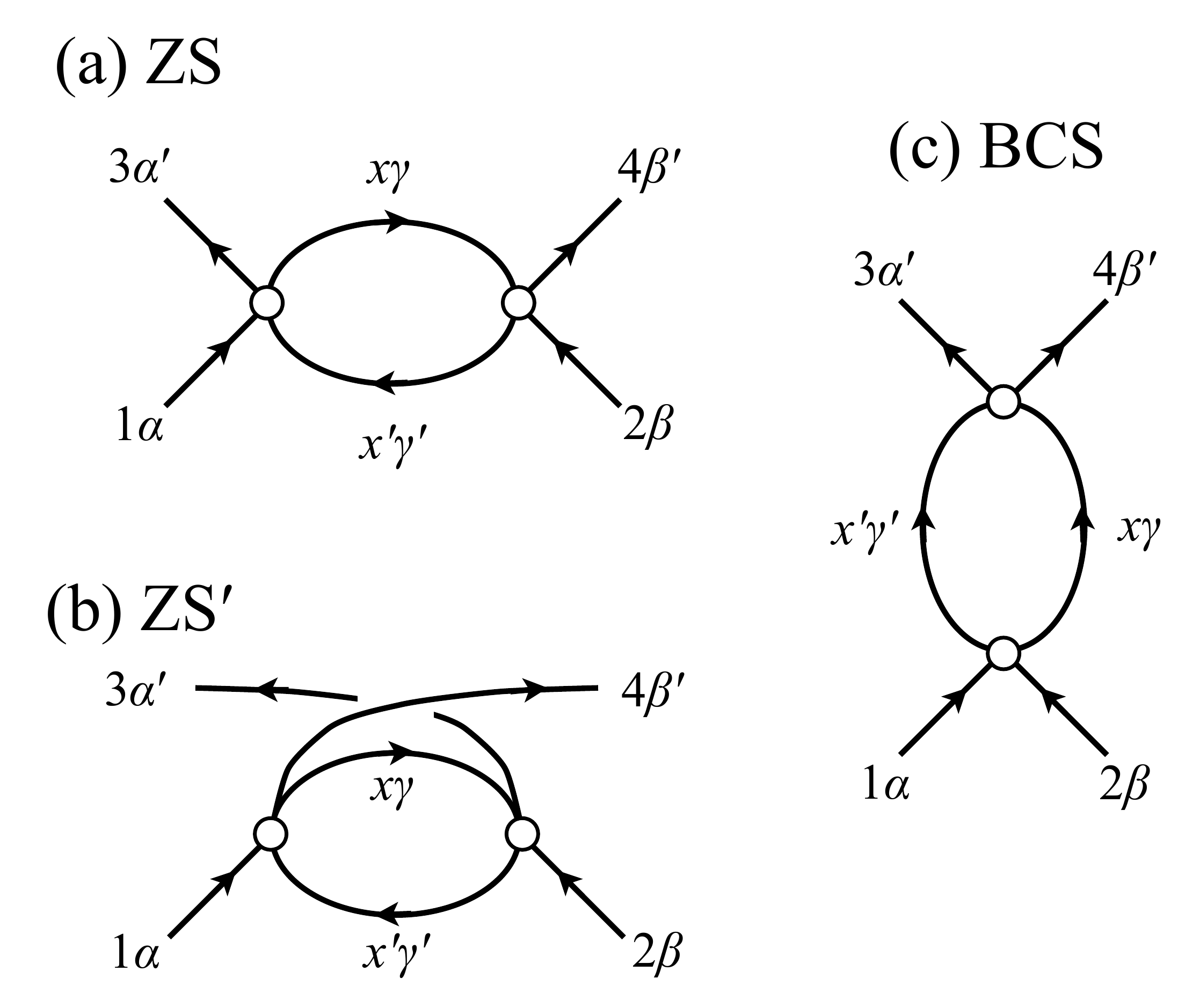}
\caption{(a) The ZS Feynman diagram:   $V_{x3,x'1}^{\gamma\alpha',\gamma'\alpha}V_{4x',2x}^{\beta'\gamma',\beta\gamma}$.  (b) The ZS' Feynman diagram:  $V_{x4,x'1}^{\gamma\beta',\gamma'\alpha}V_{3x',2x}^{\alpha'\gamma',\beta\gamma}$.  (c) The BCS Feynman diagram: $V_{43,xx'}^{\beta'\alpha',\gamma\gamma'}V_{xx',21}^{\gamma\gamma',\beta\alpha}$. }
\label{u2-diag}
\end{center}
\end{figure}

We use the nomenclature of Shankar~\cite{Shankar94} (as did Ref.~\onlinecite{Bishara09a}),
(see Fig.~\ref{u2-diag}) for the three diagrams which renormalize the two-body interaction as
\begin{equation}
\label{eqn:2-body-naive}
\delta u^{(2)}_{43;21}\equiv \mathrm{ZS} + \mathrm{ZS'} + \mathrm{BCS}
\end{equation}
where ZS/ZS' stands for ``zero-sound" and BCS stands for
``Bardeen-Cooper-Schrieffer" due to their 
similarity to the corresponding diagrams in Fermi liquid theory.
The corresponding expressions are:
\begin{eqnarray}
\mathrm{ZS} &\equiv& \int_{-\infty}^\infty\frac{d\omega_x}{2\pi}\int_{-\infty}^\infty \frac{d\omega_{x'}}{2\pi}
\sum V_{x3,x'1}^{\gamma\alpha',\gamma'\alpha}V_{4x',2x}^{\beta'\gamma',\beta\gamma}\nonumber\\
&&\times2\pi\delta(\omega_3+\omega_x-\omega_1-\omega_{x'})G_{x'}G_{x}\;, \nonumber \\
\end{eqnarray}
\begin{eqnarray}
\mathrm{ZS'} &\equiv& -\int_{-\infty}^\infty\frac{d\omega_x}{2\pi}\int_{-\infty}^\infty \frac{d\omega_{x'}}{2\pi}
\sum V_{x4,x'1}^{\gamma\beta,\gamma'\alpha}
V_{3x',2x}^{\alpha'\gamma',\beta\gamma}\nonumber\\
&&\times2\pi\delta(\omega_x+\omega_4-\omega_1-\omega_{x'})G_{x'}G_{x}\;,\nonumber\\
\end{eqnarray}
and
\begin{eqnarray}
\mathrm{BCS} &\equiv& -\frac{1}{2}\int_{-\infty}^\infty\frac{d\omega_x}{2\pi}\int_{-\infty}^\infty \frac{d\omega_{x'}}{2\pi}
\sum V_{xx',21}^{\gamma\gamma',\beta\alpha}V_{43,xx'}^{\beta'\alpha',\gamma\gamma'}\nonumber\\
&&\times2\pi\delta(\omega_{x'}+\omega_x-\omega_1-\omega_2)G_{x'}G_{x}\;,\nonumber\\
\end{eqnarray}
where the electron propagator is $G_x=1/(i\omega_x - \tilde{E}_{s_x n_x})$ 
with $\tilde{E}_{s_x n_x} = E_{s_x n_x} - \mu$.  We assume 
the chemical potential to be the energy of the partially filled Landau level, $\mu=\hbar\omega_c(N+1/2)$, and  the above $\Sigma$ 
is shorthand for the sum
\begin{eqnarray}
\label{sum}
\sum\equiv \sum_{m_x,m_{x'}=0}^\infty \sideset{}{'}\sum_{s_x,s_{x'}=0}^\infty \sideset{}{'}\sum_{n_x,n_{x'}=0}^\infty 
\sum_{\gamma,\gamma'=\downarrow,\uparrow}\;.
\end{eqnarray}
Note the primes on the sums over sub-bands ($s_x$ and $s_{x'}$) and Landau levels ($n_x$ and $n_{x'}$) indicate that 
we do not simultaneously include the conditions ($n_x=N,s_x=0$) or ($n_{x'}=N,s_{x'}=0$)--we do not want to integrate
out the $N^{\rm th}$ Landau level of the lowest sub-band.

We now turn to the evaluation of these expressions.
Consider the ZS term, for example, and integrate over $\omega_{x'}$ using the $\delta$-function:
\begin{eqnarray}
\mathrm{ZS} &=& \int_{-\infty}^\infty\frac{d\omega_x}{2\pi}
\sum
V_{x3,x'1}^{\gamma\alpha',\gamma'\alpha}V_{4x',2x}^{\beta'\gamma',\beta\gamma}\nonumber\\
&&\times \frac{1}{i(\omega_3-\omega_1+\omega_x) - 
\tilde{E}_{s_{x'} n_{x'}}}\frac{1}{i\omega_x-\tilde{E}_{s_x n_x}}
\end{eqnarray}
and we can do the remaining integral over $\omega_x$ using the identity
\begin{eqnarray}
\int_{-\infty}^\infty d\omega \frac{1}{i\omega -A}\frac{1}{i\omega-B}=2\pi\frac{\theta(\mathrm{Re}(A))-\theta(\mathrm{Re}(B))}{B-A}
\end{eqnarray}
for $\mathrm{Re}(A)\neq 0$ or $\mathrm{Re}(B)\neq 0$.  Thus,
\begin{eqnarray}
\label{eqn:ZS}
\mathrm{ZS} &=& 
\sum
V_{x3,x'1}^{\gamma\alpha',\gamma'\alpha}V_{4x',2x}^{\beta'\gamma',\beta\gamma}\nonumber \\
&&\times \frac{\theta(\tilde{E}_{s_x n_x}) - 
\theta(E_{s_{x'} n_{x'}})}{i(\omega_1-\omega_3)+\tilde{E}_{s_{x'} n_{x'}}-\tilde{E}_{s_x n_x}}\;.
\end{eqnarray}
In a similar fashion, we have
\begin{eqnarray}
\label{eqn:ZS'}
\mathrm{ZS'} &=& -
\sum V_{x4,x'1}^{\gamma\beta,\gamma'\alpha}
V_{3x',2x}^{\alpha'\gamma',\beta\gamma}\nonumber \\ &&\times \frac{\theta(\tilde{E}_{s_x n_x})-\theta(\tilde{E}_{s_{x'} n_{x'}})}
{i(\omega_1-\omega_4)+\tilde{E}_{s_{x'} n_{x'}}-\tilde{E}_{s_x n_x}}\;,
\end{eqnarray}
and
\begin{eqnarray}
\label{eqn:BCS}
\mathrm{BCS} &=& \frac{1}{2}
\sum
V_{xx',21}^{\gamma\gamma',\beta\alpha}V_{43,xx'}^{\beta'\alpha',\gamma\gamma'}\nonumber \\ &&\times
\frac{\theta(\tilde{E}_{s_x n_x})-\theta(-\tilde{E}_{s_{x'} n_{x'}})}
{i(\omega_1+\omega_2)-\tilde{E}_{s_{x'} n_{x'}} - \tilde{E}_{s_x n_x}}\;.
\end{eqnarray}
We then approximate the denominators as, for example,  
$i(\omega_1-\omega_3)+\tilde{E}_{s_{x'} n_{x'}}-\tilde{E}_{s_x n_x}\approx  \tilde{E}_{s_{x'} n_{x'}}-\tilde{E}_{s_x n_x}$
since $\hbar\omega_c$ is 
 much larger than the frequencies at which we probe 
the system. In principle, however, these effective interactions become retarded
if we do not drop the frequency dependence.
This approximation eventually yields
\begin{eqnarray}
\label{zs}
\mathrm{ZS} &\approx& 
\sum
V_{x3,x'1}^{\gamma\alpha',\gamma'\alpha}V_{4x',2x}^{\beta'\gamma',\beta\gamma}\nonumber \\ &&\times \frac{\theta(\tilde{E}_{s_x n_x}) - \theta(\tilde{E}_{s_{x' }n_{x'}})}{\tilde{E}_{s_{x' }n_{x'}}-\tilde{E}_{s_x n_x}}\;,
\end{eqnarray}
\begin{eqnarray}
\label{zsp}
\mathrm{ZS'} &\approx&  -
\sum V_{x4,x'1}^{\gamma\beta,\gamma'\alpha}
V_{3x',2x}^{\alpha'\gamma',\beta\gamma}\nonumber \\ &&\times \frac{\theta(\tilde{E}_{s_x n_x})-\theta(\tilde{E}_{s_{x'} n_{x'}})}
{\tilde{E}_{s_{x'} n_{x'}}-\tilde{E}_{s_x n_x}}\;,
\end{eqnarray}
and
\begin{eqnarray}
\label{bcs}
\mathrm{BCS} &\approx& \frac{1}{2}
\sum
V_{xx',21}^{\gamma\gamma',\beta\alpha}V_{43,xx'}^{\beta'\alpha',\gamma\gamma'}\nonumber\\&&\times
\frac{\theta(\tilde{E}_{s_x n_x})-\theta(-\tilde{E}_{s_{x'} n_{x'}})}
{-\tilde{E}_{s_{x'} n_{x'}}-\tilde{E}_{s_x n_x}}\;.
\end{eqnarray}

We can extract some physics from these formulas
without calculating any numbers.  
Each interaction matrix element ($V_{x4,x'1}^{\gamma\beta,\gamma'\alpha}$
from the ZS diagram, for example) comes 
with a factor of $e^2/\epsilon \ell_0$ according to
Eq. ~(\ref{eqn:V4321-def}). Therefore, once a factor of $\hbar \omega_c$
is pulled out of the denominator, the ZS, ZS', and BCS contributions
to $u^{(2)}_{43;21}$ are proportional to $\kappa (e^2/\epsilon \ell_0)$, as expected. In the lowest Landau level ($N=0$),
the ZS and ZS' terms vanish
since there cannot be any hole excitations in the internal legs
(the $x'\gamma'$ leg) since there are no filled Landau levels below in which to have virtual hole excitations.
Therefore, only the BCS diagram renormalizes the two-body interactions in the lowest Landau level--incidentally, this is 
\textit{not} the case for graphene since there all Landau levels $N=0$ are filled in which it is possible to excite virtual  holes.

\begin{figure}[htbp]
\begin{center}
\includegraphics[width=5.5cm,angle=0]{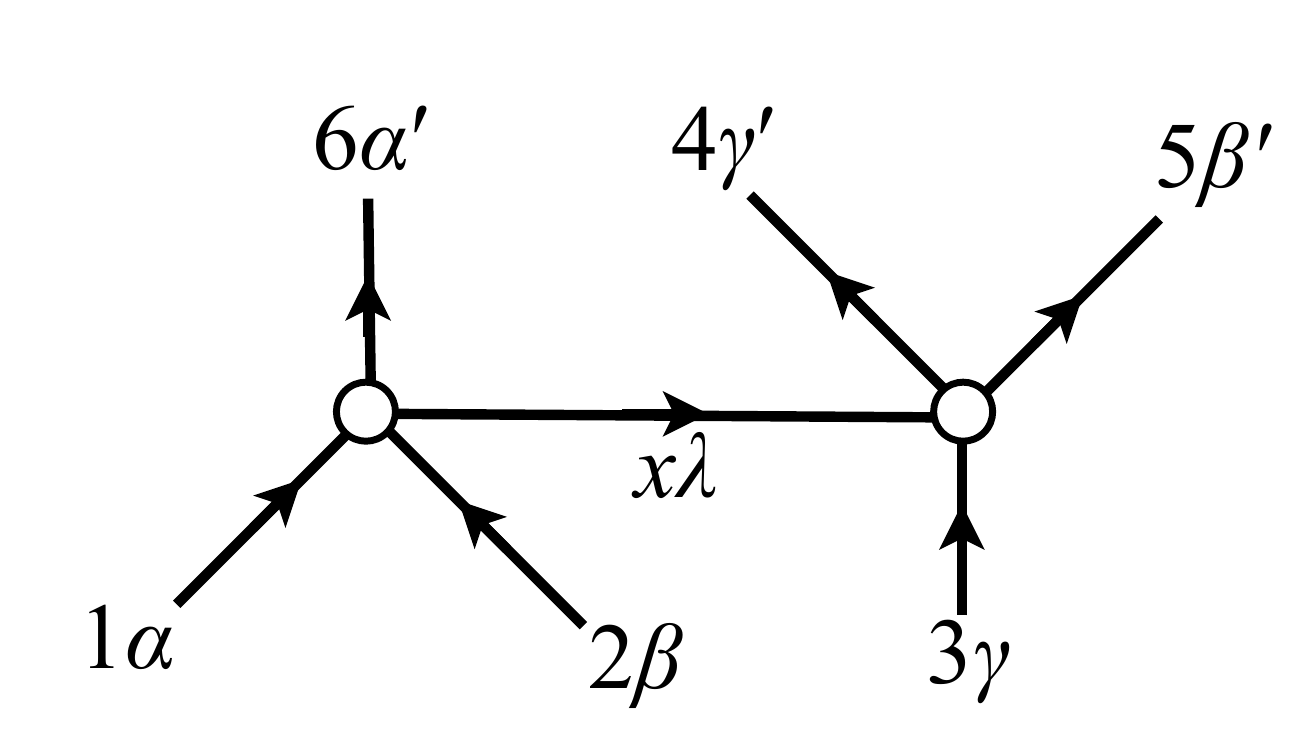}
\caption{One of the nine three-body diagrams.}
\label{u3-diag}
\end{center}
\end{figure}

We now turn to the three-body interaction $u^{(3)}_{654;321}$.
At order $\kappa$ there is a three-body term that is 
generated in our expansion of the action.
Figure~\ref{u3-diag} shows one of the nine diagrams that contribute to the 
three-body term. The expression for the sum of all nine diagrams can be written compactly~\cite{Bishara09a,bishara-thesis} as
\begin{eqnarray}
\label{3body}
 \sideset{}{'}\sum_{n_x,s_x=0}^{\infty} \sum_{m_x=0}^\infty\sum_{\gamma=\uparrow,\downarrow}
\sum_\mathrm{cyc.\;perm.} \frac{{V}_{6x,21}^{\alpha'\lambda,\beta\alpha}{V}_{54,x3}^{\beta'\gamma',\lambda\gamma}}
{i\omega_x-\tilde{E}_{s_x,n_x}}\;.
\end{eqnarray}
At low-frequencies, this leads to a three-body term:
\begin{eqnarray}
\label{3bodylow}
u^{(3)}_{654:321}&=& - \sideset{}{'}\sum_{n_x,s_x=0}^{\infty} \sum_{m_x=0}^\infty\sum_{\gamma=\uparrow,\downarrow}\nonumber\\
&&\times \sum_\mathrm{cyc.\;perm.} \frac{{V}_{6x,21}^{\alpha'\lambda,\beta\alpha}{V}_{54,x3}^{\beta'\gamma',\lambda\gamma}}{\tilde{E}_{s_x,n_x}}
\end{eqnarray}
where we sum over all cyclic permutations of the indices labeled with $(1,2,3)$ and $(4,5,6)$.  The prime on the summation 
over $n_x$ and $s_x$ indicates that we do not include cases where $s_x=0$ and $n_x=N$ simultaneously.

\subsection{Normal-Ordering the Three-Body Interaction}
\label{subsec:normal-ordering}

There is one important subtlety that we now discuss.
When we compute the three-body term, just discussed, that is induced
at order $\kappa$, we evaluate diagrams that correspond
to the operator expression:
\begin{equation}
\frac{1}{2!}\left[
\contraction{c_6^\dagger}{c}{_x^\dagger \,c^{}_2 c^{}_1 \,\, c_5^\dagger c_4^\dagger\,}{c}
c_6^\dagger c_x^\dagger \,c^{}_2 c^{}_1 \,\,c_5^\dagger c_4^\dagger \,c^{}_x c^{}_3
+
\contraction{c_5^\dagger c_4^\dagger \,c^{}_3}{c}{^{}_x\,\,}{c}
c_5^\dagger c_4^\dagger \,c^{}_3\, c^{}_x \,\,c_x^\dagger c_6^\dagger \,c_2 \,c_1 \right]
\end{equation}
Here, we have subsumed spin, Landau level, sub-band, and orbital indices into a single index for brevity. In the functional integral, we do not need to be careful about the order of Grassmann variables, apart from keeping track of signs. However, when we pass from the functional integral to the operator formalism, we must be careful about the order of these operators. The order of these operators is determined by a point-splitting regularization:
\begin{eqnarray}
\int d\omega e^{i\omega 0^+}c^\dagger(\omega)c(\omega)&=&c^\dagger(t)c(t)\nonumber\\
\int d\omega e^{i\omega 0^-}c^\dagger(\omega)c(\omega)&=&c(t)c^\dagger(t)\;.
\end{eqnarray}
In other words, the order of operators at the same time is determined by whether the contour must be closed in the upper- or lower-half-plane. In the case of our three- body interaction, which is actually a retarded interaction, as in Eq.~(\ref{3body}), this is determined by the sign of $\tilde{E}_{s_x,n_x}$.  When we make the approximation of neglecting $\omega_x$ in 
the denominator of Eq. (\ref{3body}) to pass to the low-energy limit in Eq. (\ref{3bodylow}), we can drop 
the $\omega_x$-dependence, \emph{except} that we must remember the operator ordering that it imposes. Therefore, the three-body term that we have computed is not really
\begin{eqnarray}
-\sum \frac{{V}_{6x,21}^{\alpha'\lambda,\beta\alpha}{V}_{54,x3}^{\beta'\gamma',\lambda\gamma}}{\tilde{E}_{s_x,n_x}}
c^\dagger_6 c^\dagger_5 c^\dagger_4 c_3 c_2 c_1
\end{eqnarray}
but is, instead,
\begin{multline}
-\frac{1}{2}\sum \frac{{V}_{6x,21}^{\alpha'\lambda,\beta\alpha}{V}_{54,x3}^{\beta'\gamma',\lambda\gamma}}{\tilde{E}_{s_x,n_x}}\times\\
\left[
\theta(\tilde{E}_{s_x,n_x}) c_6^\dagger  c_5^\dagger c_4^\dagger \, c^{}_3 c^{}_2 c^{}_1
+ \theta(-\tilde{E}_{s_x,n_x}) c_6^\dagger c^{}_2 c^{}_1 c_5^\dagger c_4^\dagger \, c^{}_3
\right.\\
\left. + \theta(-\tilde{E}_{s_x,n_x}) c_6^\dagger c_5^\dagger c_4^\dagger \, c^{}_3 \,
c^{}_2 c^{}_1
- \theta(\tilde{E}_{s_x,n_x}) c_5^\dagger c_4^\dagger \, c^{}_3 \,
c_6^\dagger c^{}_2 c^{}_1\right]\;.
\end{multline}
After we normal-order, this can be re-written as:
\begin{eqnarray}
\label{new2body}
&&\hspace{-1cm}\frac{1}{3!}u^{(3)}_{654;321}c^\dagger_6 c^\dagger_5 c^\dagger_4 c_3 c_2 c_1\nonumber\\
&&+\sum\frac{{V}_{x3,x_0 1}^{\gamma\alpha',\gamma'\alpha}{V}_{4x_0,2x}^{\beta'\gamma',\beta\gamma}}{\tilde{E}_{s_x,n_x}}\theta(-\tilde{E}_{s_x,n_x})c^\dagger_4 c^\dagger_3 c_2 c_1\nonumber\\
&&-\sum\frac{{V}_{x4,x_0 1}^{\gamma\beta',\gamma'\alpha}{V}_{3x_0,2x}^{\alpha'\gamma',\beta\gamma}}{\tilde{E}_{s_x,n_x}}\theta(-\tilde{E}_{s_x,n_x})c^\dagger_4 c^\dagger_3 c_2 c_1\nonumber\\
&&-\frac{1}{2}\sum\frac{{V}_{xx_0,21}^{\gamma\gamma',\beta\alpha}{V}_{43,xx_0}^{\beta'\alpha',\gamma\gamma'}}{\tilde{E}_{s_x,n_x}}\theta(\tilde{E}_{s_x,n_x})c^\dagger_4 c^\dagger_3 c_2 c_1\;,
\end{eqnarray}
where the $x_0$ indicates that $n_x=N$.
Therefore, the renormalization of the two-body interaction takes the form
\begin{eqnarray}
\label{eqn:2bodyfinal}
\frac{1}{2!}\left[\mathrm{ZS}+\mathrm{ZS'}+\mathrm{BCS}\right]c^\dagger_4 c^\dagger_3 c_2 c_1 +\frac{1}{2!}\delta\tilde{u}_{43;21}^{(2)},
\end{eqnarray}
where $\frac{1}{2!}\delta \tilde{u}^{(2)}_{43;21}$ is the sum of the last three terms in Eq.~(\ref{new2body}).  Note that  Eq.~(\ref{new2body}) contains the operators 
so when comparing the last three terms of Eq.~(\ref{new2body}) with Eqs.~(\ref{zs}),~(\ref{zsp}), and~(\ref{bcs}), which do not contain the operators, there is a factor of 2 that multiplies the 
former compared to the latter.  Also note that, in the $N=0$ Landau level, 
the first and second two-body terms in Eq.~(\ref{new2body}) vanish and only the third term contributes.

There is a simple check that shows that $\delta\tilde{u}^{(2)}_{43;21}$ is necessary.  Suppose that we compute two-particle scattering to second-order 
in the Coulomb interaction.  We can do this either in the full theory (with all Landau levels kept) 
or in the low-energy effective theory of the $N^{\rm th}$ Landau 
level.  The answers must be the same either way.  With the $\delta\tilde{u}^{(2)}_{43;21}$ above, this is the case.  There is a simple physical 
interpretation for the renormalization of the two-body term that results from normal-ordering the three-body term.  We can think of this term 
as arising by taking a three-body vertex and connecting one of the incoming lines to one of the out-going lines in Fig.~\ref{u3-diag}.  If the two lines are 
connected to the same vertex (for example, 1 and 6 in Fig.~\ref{u3-diag}) then we have a tadpole diagram, which vanishes.  (Two such lines should not 
have been connected anyway since the operators at a single vertex are normal ordered.)  If the two lines are connected to different vertices, then the resulting 
diagram is of the form of the ZS, ZS', or BCS diagrams (see Fig.~\ref{u2-diag}), but one of the internal lines remains in the $N^{\rm th}$ Landau level, rather 
than one of the higher or lower Landau levels that is being integrated out.  In other words, this contribution accounts for processes such 
as the following:  (i) two electrons in the $N^{\rm th}$ Landau level interact, (ii) one of them is excited virtually to a higher Landau level while the other 
remains in the $N^{\rm th}$ Landau level, and (iii) they interact again and the excited electron falls back to the $N^{\rm th}$ Landau level.

One might worry that the result of such a process will depend on the couplings within the $N^{\rm th}$ Landau level and its filling factor.  The 
two-body terms that we have generated in this section derive from normal-ordering the three-body terms, so the dynamics within the $N^{\rm th}$ Landau level
does not enter into their computation.  These dynamics (and the filling of the 
$N^{\rm th}$ Landau level) enters only when we attempt to solve the resulting effective Hamiltonian for the $N^{\rm th}$ Landau levels.

Another way to think about this is to recall that we have \emph{not} integrated out states within the $N^{\rm th}$ Landau level.  Within a 
Wilsonian renormalization group scheme, which is essentially what we have adopted, all internal lines in diagrams are at high energies and 
all external lines are at low energies because we integrate out $N'\neq N$ Landau levels but do not do anything to states in the $N^{\rm th}$ Landau 
level, which will be dealt with later (by, for instance, exact diagonalization).  In our case, this means that all internal lines are in the $N'\neq N$ Landau 
levels while all external lines are in the $N^{\rm th}$ Landau level.  In such an approach, a diagram with $k$ external legs in which $p$ internal lines 
are in the $N'\neq N$ Landau levels while $q$ are in the $N^{\rm th}$ Landau level arises in the following way.  When we integrate out the 
$N'\neq N$ Landau levels, we will generate a vertex with $k+2q$ legs.  One of the terms contributing to this vertex will have $p$ internal lines.  If 
we were to solve the problem in the $N^{\rm th}$ Landau level perturbatively, then there would be a diagram in which we took this $(k+2q)$-leg vertex 
and connected $q$ incoming lines to $q$ outgoing lines.  Therefore, this physics is present in the $(k+2q)$-leg vertex.  In our case, this means that 
two-body processes in which the intermediate state has one electron in the $N^{\rm th}$ Landau level and one in an $N'\neq N$ Landau level 
are present in our three-body vertex.  However, to correctly account for them, it is crucial to order the operators in the three-body vertex 
correctly.  

Note that in Fermi liquid theory for electrons with no external magnetic field ($B=0$) we never have to consider the type of 
virtual process which gives this contribution because momentum conservation does not allow processes in which one electron is scattered 
to momenta far from the Fermi surface while the other electron stays near the Fermi surface.  In low Landau levels, by contrast, there is 
no conservation law precluding such processes and they are not only present but, in fact, give a substantial contribution 
to the renormalization of the two-body interaction.

\subsection{Effective Hamiltonian}
\label{subsec:effective-ham-gaas}

Once we have these new interactions, $u^{(2)}_{43;21}$ and $u^{(3)}_{654;321}$,
we can calculate the Haldane pseudopotentials 
and their three-body generalizations~\cite{Simon07c}. 
We use $V^{(2)}_{MS}(N)$ to denote the bare two-body
pseudopotential for two electrons residing in the $N^{\rm th}$ Landau level with
relative angular momentum $M$ and total spin $S$.
The $\mathcal{O}(\kappa)$ correction (including the contribution
discussed in the previous subsection) to this pseudopotential
is denoted by $\delta V^{(2)}_{MS}(N)$.
The three-body pseudopotential for three electrons
with relative angular momentum $M$ (defined as the degree of the
relative wavefunction)\cite{Simon07c} and total spin $S$ is
denoted by $V^{(3)}_{MS}(N)$:  
\begin{multline}
\label{v2m}
V^{(2)}_{MS}(N)=\sum \langle MS|m_3\alpha',m_4\beta'\rangle \langle m_1\alpha,m_2\beta|MS\rangle  \\
\times V_{43;21}
\end{multline}
\begin{multline}
\label{dv2m}
\delta V^{(2)}_{MS}(N)=\sum \langle MS|m_3\alpha',m_4\beta'\rangle \langle m_1\alpha,m_2\beta|MS\rangle\\
\times \delta u^{(2)}_{43;21}
\end{multline}
\begin{multline}
\label{v3m}
V^{(3)}_{MS}(N)=\sum \langle MS|m_4\gamma',m_5\beta',m_6\alpha'\rangle
\, \times\ \\
\langle m_1\alpha,m_2\beta,m_3\gamma|MS\rangle u^{(3)}_{654;321}
\end{multline}
where $\sum$ indicates a sum over all $m_i$ and primed spin variables.  

We will drop the spin  indices from the pseudopotentials we are considering in this work and write $V^{(2)}_{M}(N)\equiv V^{(2)}_{MS}(N)$,
$V^{(3)}_{M}(N)\equiv V^{(3)}_{MS}(N)$ where 
the spin will be $S=1/2$ for $M=1$ and 2 and $S=3/2$ for $M\geq3$ for the three-body case, while for 
the two-body case, the spin will be $S=1$(0) for spin polarized (unpolarized) electrons, i.e., $M=$odd (even).
Our Hamiltonian is appropriate for all filling factors.  
Therefore, we calculate even and odd $M$ for the two-body terms 
in addition to $M=1$, 2, 3, 5, 6, 7, and 8 for the three-body terms.  It should be clear what the total spin is for those particular $M$'s.
As shown in Section~\ref{pseudoVms}, the numerical value for $V^{(3)}_{8}(0)$,
$V^{(3)}_{8}(1)$ are very small, so we ignore $M>8$.
This allows us to avoid the potential complication
of pseudopotential matrices, instead of pseudopotential numbers, 
that arise for total relative angular momentum $M\geq9$ due to the multiple three-particle  wavefunctions~\cite{Simon07c}.

Starting with Eq.~(\ref{gab}), we have chosen
to calculate the effective interactions and the resulting
pseudopotentials in the planar geometry, rather than in
the spherical geometry.  For the two-body pseudopotentials,
we are only reporting the {\it corrections} due 
to Landau level and sub-band mixing and
one must consider these corrections as corrections to the {\it planar} bare 
pseudopotentials.  
In order to calculate them in the spherical geometry,
we would have to recalculate the corrections for each system size, which
is extremely cumbersome and, presumably unnecessary (when we
extrapolate to the thermodynamic limit).
The spherical pseudopotentials  approach the planar pseudopotentials
in the thermodynamic  limit of a large sphere and it has been
argued~\cite{Peterson08,Peterson08b} that  using the planar pseudopotentials in finite 
systems better approximate the infinite system.

With these new effective pseudopotentials, we construct a
Hamiltonian for the FQHE in the $N^{\rm th}$ Landau level
 which includes Landau level and sub-band mixing 
effects perturbatively to first order in $\kappa$, 
\begin{multline}
\label{hbn}
\hat{H}_{\rm eff}(\kappa,d/\ell_0) = \hat{V}_0^{(2)} (d/\ell_0) + \kappa \hat{V}_1^{(2)} (d/\ell_0)\\
+ \kappa \hat{V}_1^{(3)} (d/\ell_0)
\end{multline}
where 
\begin{multline}
\hat{V}_0^{(2)}(d/\ell_0) =\sum_{i<j}V(|\mathbf{r}_i-\mathbf{r}_j|)\\
= \sum_M V_{M}^{(2)}(N,d/\ell_0)\sum_{i<j}\hat{P}_{ij}(M)
\end{multline}
is the bare pseudopotentials with $V(|\mathbf{r}_i-\mathbf{r}_j|)$ being the usual finite-thickness augmented Coulomb interaction 
(see Refs.~\onlinecite{Peterson08,Peterson08b}); it is 
a function of the quantum-well thickness $d/\ell_0$.
In this expression, $\hat{P}_{ij}(M)$ is an operator that projects the pair of electrons $(i,j)$
onto a two-body state of relative angular momentum $M$.
Similarly,
\begin{equation}
\hat{V}_1^{(2)}(d/\ell_0) = \sum_M \delta V_{M}^{(2)}(N,d/\ell_0)\sum_{i<j}\hat{P}_{ij}(M)
\end{equation}
and
\begin{equation}
\hat{V}_1^{(3)}(d/\ell_0)=\sum_L V_{M}^{(3)}(N,d/\ell_0)\sum_{i<j<k}\hat{P}_{ijk}(M)
\end{equation}
where $\hat{P}_{ijk}(M)$ is a  projection operator that projects onto triplets
of electrons with relative angular momentum $M$.  The subscripts $0$ and $1$ in Eq.~(\ref{hbn})
indicate the bare interaction and the first order corrections, respectively.  We also briefly comment 
about the projection operators:
what is often not explicitly written is that the $m$-body projection operators
$\hat{P}_{ij\ldots m}(M)$ have normalization set by the property
that their eigenvalues must be $0$ or $1$.

Since every Landau level has the same number of states,
we project the $N^{\rm th}$ Landau level to the $N=0$ Landau
level. The difference between the $N^{\rm th}$ Landau level and the zeroth
is reflected in the values of the pseudopotentials. However, 
from the point of view of any calculations,
our effective Hamiltonian, Eq.~(\ref{hbn}), is a Hamiltonian
for electrons in the lowest Landau level that simulates
the Hamiltonian for electrons in the $N^{\text{th}}$ Landau level.

\section{THE EFFECTIVE TWO- AND THREE-BODY PSEUDOPOTENTIALS for $\mbox{GaAs}$}
\label{pseudoVms}

We now discuss the numerical values of the two- and three-body
pseudopotentials in the presence of Landau-level and sub-band mixing appropriate to GaAs (i.e. for parabolic $B=0$ bands).  
For the two-body case, we calculate  Eq.~(\ref{eqn:2bodyfinal}). The result
is a function of single-particle angular momenta
$m_1$, $m_2$, $m_3$, and $m_4$,  denoted
by $\delta u^{(2)}_{43:21}$, which depends on the Landau level
index $N$ of the partially-filled Landau level which we retain.  All other Landau levels
are either completely empty or completely full and are integrated out.
Then, $\delta u^{(2)}_{43:21}$  is substituted into Eq.~(\ref{dv2m}), yielding 
the two-body pseudopotential correction $\delta V^{(2)}_{M}(N)$.  
For the three-body term we calculate $u^{(3)}_{654;321}$ (Eq.~(\ref{3body}))
which is a function of angular momenta $m_6$, $m_5$, $m_4$, $m_3$, $m_2$, and $m_1$.
The result, which depends on the index $N$ of the partially-filled Landau level which
we retain, is substituted into Eq.~(\ref{v3m}), yielding the three-body pseudopotential
$V^{(3)}_{M}(N)$.

However, Eqs.~(\ref{eqn:2bodyfinal}), and~(\ref{3body})
all contain infinite sums over  Landau levels and sub-bands and
none of these sums can be done analytically. Therefore, they must
be truncated and the resulting finite sums must be computed numerically.
The infinite sum is the limit of the finite sums
as the truncation is taken to infinity.
We calculate our final results [$\delta V^{(2)}_{M}(N,{d/\ell_0})$ and
$V^{(3)}_{M}(N,{d/\ell_0})$] as a 
function of the truncation and determine the convergence to the infinite sum by fitting to some
chosen general common function. 

\begin{table}[h]
\caption{We list the numerical values of $\delta V^{(2)}_{M}(N,d/\ell_0)$ for $d/\ell_0=0$, 1, 2, 3, 4 and $N=0,1$.  Values for $d/\ell_0=0$ 
were given previously in Ref.~\onlinecite{Bishara09a} for $M=1,3$ but due to the new two-body correction 
coming from properly normal-ordering the three-body interaction the values have been changed~\cite{footnote-sodemann}.  All energies 
are given in units of $e^2/\epsilon\ell_0$.}
\begin{ruledtabular}
\label{table:2-body-pseudo}
\begin{tabular}{lrrrrr}
 & $d/\ell_0=0 $ &  1  & 2 & 3 & 4 \\ \hline
$\delta V_{0}^{(2)}(0,{d/\ell_0})$ & -0.3422 & -0.1963 & -0.1281 & -0.0900 & -0.0665 \\
$\delta V_{1}^{(2)}(0,{d/\ell_0})$ & -0.0328 & -0.0300 & -0.0254 & -0.0211 & -0.0175 \\
$\delta V_{2}^{(2)}(0,{d/\ell_0})$ & -0.0112 & -0.0108 & -0.0098 & -0.0088 & -0.0077 \\
$\delta V_{3}^{(2)}(0,{d/\ell_0})$ & -0.0055 & -0.0054 & -0.0051 & -0.0047 & -0.0043 \\
$\delta V_{4}^{(2)}(0,{d/\ell_0})$ & -0.0033 & -0.0032 & -0.0031 & -0.0029 & -0.0027 \\
$\delta V_{5}^{(2)}(0,{d/\ell_0})$ & -0.0022 & -0.0022 & -0.0021 & -0.0020 & -0.0019 \\
$\delta V_{6}^{(2)}(0,{d/\ell_0})$ & -0.0016 & -0.0015 & -0.0015 & -0.0014 & -0.0014 \\
$\delta V_{7}^{(2)}(0,{d/\ell_0})$ & -0.0012 & -0.0011 & -0.0011 & -0.0011 & -0.0010 \\
$\delta V_{8}^{(2)}(0,{d/\ell_0})$ & -0.0009 & -0.0009 & -0.0009 & -0.0009 & -0.0008 \\
$\delta V_{9}^{(2)}(0,{d/\ell_0})$ & -0.0007 & -0.0007 & -0.0007 & -0.0007 & -0.0007 \\
\hline
$\delta V_{0}^{(2)}(1,{d/\ell_0})$ & -0.3816 & -0.3184 & -0.2696 & -0.2307 & -0.1997 \\
$\delta V_{1}^{(2)}(1,{d/\ell_0})$ & -0.2143 & -0.2020 & -0.1815 & -0.1617 & -0.1442 \\
$\delta V_{2}^{(2)}(1,{d/\ell_0})$ & -0.1787 & -0.1456 & -0.1296 & -0.1172 & -0.1067 \\
$\delta V_{3}^{(2)}(1,{d/\ell_0})$ & -0.1039 & -0.0986 & -0.0927 & -0.0868 & -0.0812 \\
$\delta V_{4}^{(2)}(1,{d/\ell_0})$ & -0.0789 & -0.0729 & -0.0694 & -0.0664 & -0.0635 \\
$\delta V_{5}^{(2)}(1,{d/\ell_0})$ & -0.0353 & -0.0423 & -0.0458 & -0.0473 & -0.0476 \\
$\delta V_{6}^{(2)}(1,{d/\ell_0})$ & -0.0258 & -0.0296 & -0.0334 & -0.0361 & -0.0376 \\
$\delta V_{7}^{(2)}(1,{d/\ell_0})$ & -0.0115 & -0.0181 & -0.0234 & -0.0272 & -0.0297 \\
$\delta V_{8}^{(2)}(1,{d/\ell_0})$ & -0.0073 & -0.0123 & -0.0175 & -0.0216 & -0.0245 \\
$\delta V_{9}^{(2)}(1,{d/\ell_0})$ & -0.0023 & -0.0079 & -0.0133 & -0.0176 & -0.0209 \\
\end{tabular}
\end{ruledtabular}
\end{table}

\begin{table}[h]
\caption{We list the numerical values of the three-body pseudopotentials
$V^{(3)}_{M}(N,{d/\ell_0})$
as a function of $M$ for $d/\ell_0 = 0$, 1, 2, 3, and 4 and $N=0,1$. The values for $d/\ell_0=0$ 
were given previously in Ref.~\onlinecite{Bishara09a} where again a 
few values have been changed very slightly due 
to a more careful extrapolation. All energies 
are given in units of $e^2/\epsilon\ell_0$.}
\begin{ruledtabular}
\label{table:3-body-pseudo}
\begin{tabular}{lrrrrr}
  & $d/\ell_0=0$ &  1 & 2 & 3 & 4  \\ \hline
$ V_{1}^{(3)}(0,{d/\ell_0})$ & -0.0345 & -0.0400 & -0.0364 & -0.0312 & -0.0263  \\ 
$ V_{2}^{(3)}(0,{d/\ell_0})$ & -0.0540 & -0.0410 & -0.0324 & -0.0262 & -0.0216 \\ 
$ V_{3}^{(3)}(0,{d/\ell_0})$ & -0.0181 & -0.0173 & -0.0156 & -0.0138 & -0.0120  \\
$ V_{5}^{(3)}(0,{d/\ell_0})$ &  0.0033 & 0.0026 & 0.0015 & 0.0006 & -0.0000 \\ 
$ V_{6}^{(3)}(0,{d/\ell_0})$ & -0.0107 & -0.0102 & -0.0093 & -0.0083 & -0.0073 \\
$ V_{7}^{(3)}(0,{d/\ell_0})$ &  0.0059 & 0.0054 & 0.0043 & 0.0033 & 0.0025 \\ 
$ V_{8}^{(3)}(0,{d/\ell_0})$ & -0.0047 & -0.0045 & -0.0041 & -0.0037 & -0.0033 \\
\hline
$ V_{1}^{(3)}(1,{d/\ell_0})$ &  -0.0319 & -0.0280 & -0.0232 & -0.0192 & -0.0160 \\
$ V_{2}^{(3)}(1,{d/\ell_0})$ & -0.0305 & -0.0223 & -0.0174 & -0.0142 & -0.0117 \\ 
$ V_{3}^{(3)}(1,{d/\ell_0})$ & -0.0147 & -0.0136 & -0.0118 & -0.0101 & -0.0087 \\ 
$ V_{5}^{(3)}(1,{d/\ell_0})$ & -0.0054 & -0.0051 & -0.0047 & -0.0042 & -0.0038 \\ 
$ V_{6}^{(3)}(1,{d/\ell_0})$ & -0.0099 & -0.0093 & -0.0082 & -0.0071 & -0.0061 \\ 
$ V_{7}^{(3)}(1,{d/\ell_0})$ &   0.0005 & 0.0002 & -0.0001 & -0.0004 & -0.0006 \\ 
$ V_{8}^{(3)}(1,{d/\ell_0})$  & -0.0009 & -0.0013 & -0.0016 & -0.0017 & -0.0018  \\ 
\end{tabular}
\end{ruledtabular}
\end{table}

\begin{figure*}[t]
\begin{center}
\includegraphics[width=8.cm]{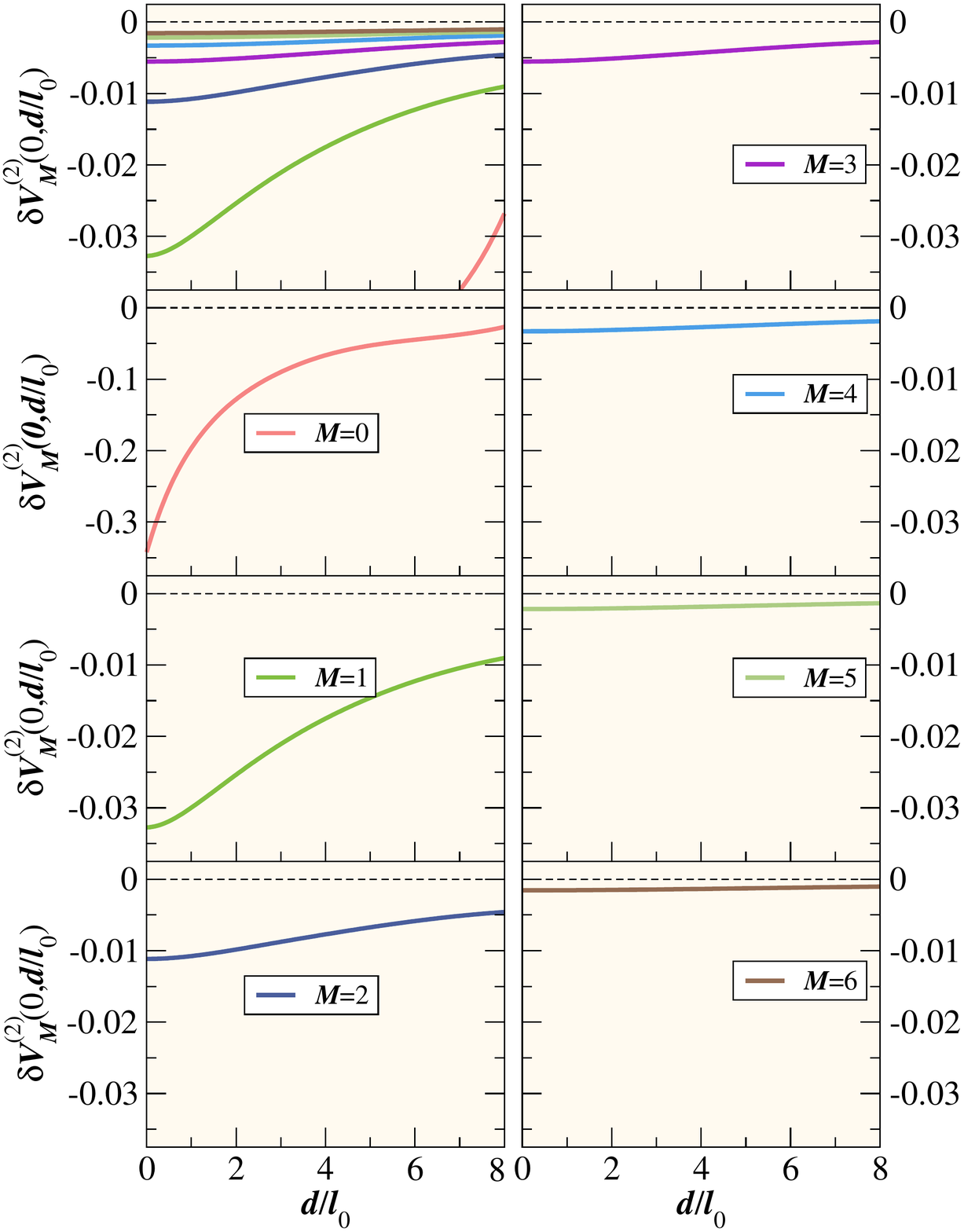}
\includegraphics[width=8.cm]{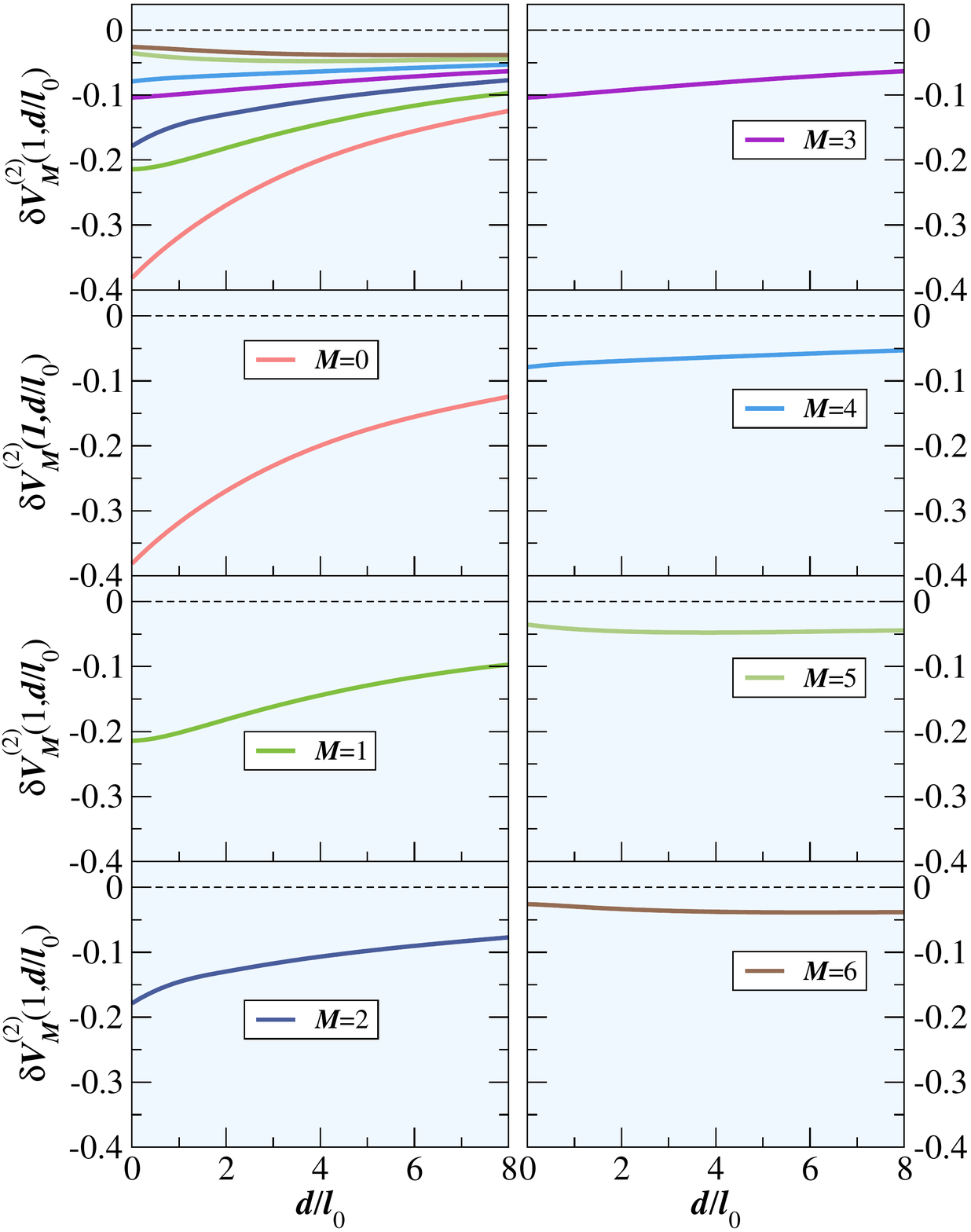}
\caption{(Color online) The Landau level mixing induced corrections to the two-body pseudopotentials for $M=1$-6 in the lowest ($N=0$, left figures) and 
second ($N=1$, right figures) Landau levels.   The top left plot in both panels shows all $\delta V^{(2)}_M(N,d/\ell_0)$ versus $d/\ell_0$ on the same plot for an easier comparison.  Note that for the lowest Landau level ($N=0)$ the $M=0$ plot is on a different vertical scale than all the others since 
its value is so much larger.  For the $N=0$ Landau level, the 
corrections are less than 10\% (for all $d/\ell_0$) of the bare pseudopotential value--discounting $M=0$ which is nearly 30\% (for $d/\ell_0 = 0$) 
the bare pseudopotential.  For the 
$N=1$ the corrections are much larger.  For $M=0$ the correction is nearly 60\% and remains near 50\% for $M<3$.  Of course, for 
both Landau levels the corrections are mitigated by finite width (increasing $d/\ell_0$). All energies 
are given in units of $e^2/\epsilon\ell_0$.}
\label{plot_vms}
\end{center}
\end{figure*}   

\begin{figure}[]
\begin{center}
\includegraphics[width=8.5cm,angle=0]{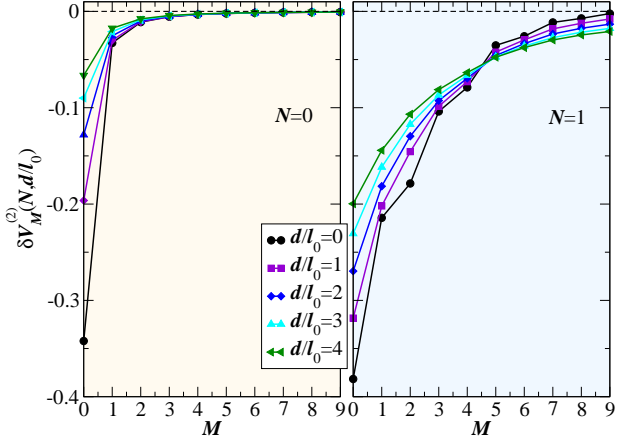}
\caption{(Color online) $\delta V^{(2)}_{M}(N,{d/\ell_0})$ as a function of $M$ for $d/\ell_0=0$, 1, 2, 3 and 4.  The left panel corresponds 
to the lowest Landau level ($N=0$) while the right panel corresponds to the first excited Landau level ($N=1$).  As in the case of the bare pseudopotentials, the 
corrections to the two-body pseudopotentials decrease with increasing $M$. All energies 
are given in units of $e^2/\epsilon\ell_0$.}
\label{plot_v2_2}
\end{center}
\end{figure}

\begin{figure}[t!]
\begin{center}
\includegraphics[width=8.5cm,angle=0]{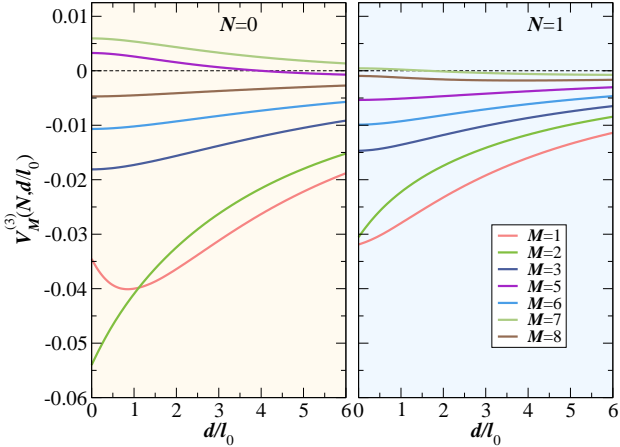}
\caption{(Color online) The three-body
pseudopotentials $V^{(3)}_{M}(N,{d/\ell_0})$
as a function of $d/\ell_0$ for the lowest Landau level (left panel) and 
the first excited Landau level (right panel).   Note that these are numerically small compared to the bare two-body 
pseudopotentials. All energies 
are given in units of $e^2/\epsilon\ell_0$.}
\label{plot_3v_1}
\end{center}
\end{figure}

\begin{figure}[t!]
\begin{center}
\includegraphics[width=8.5cm,angle=0]{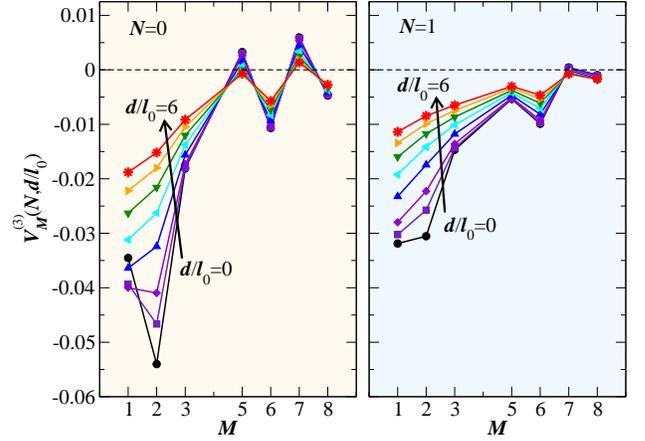}
\caption{(Color online) $V^{(3)}_{M}(N,{d/\ell_0})$ as a function of $M$ for $d/\ell_0=0$, 0.5, 1-6 
for the lowest Landau level (left panel) and the first excited Landau level (right panel).  The magnitude decreases with $M$ 
except for the non-trivial oscillatory behavior for $M\geq 5$.  All energies 
are given in units of $e^2/\epsilon\ell_0$.}
\label{plot_3v_2}
\end{center}
\end{figure}

\begin{figure*}[t!]
\begin{center}
\includegraphics[width=14.cm,angle=0]{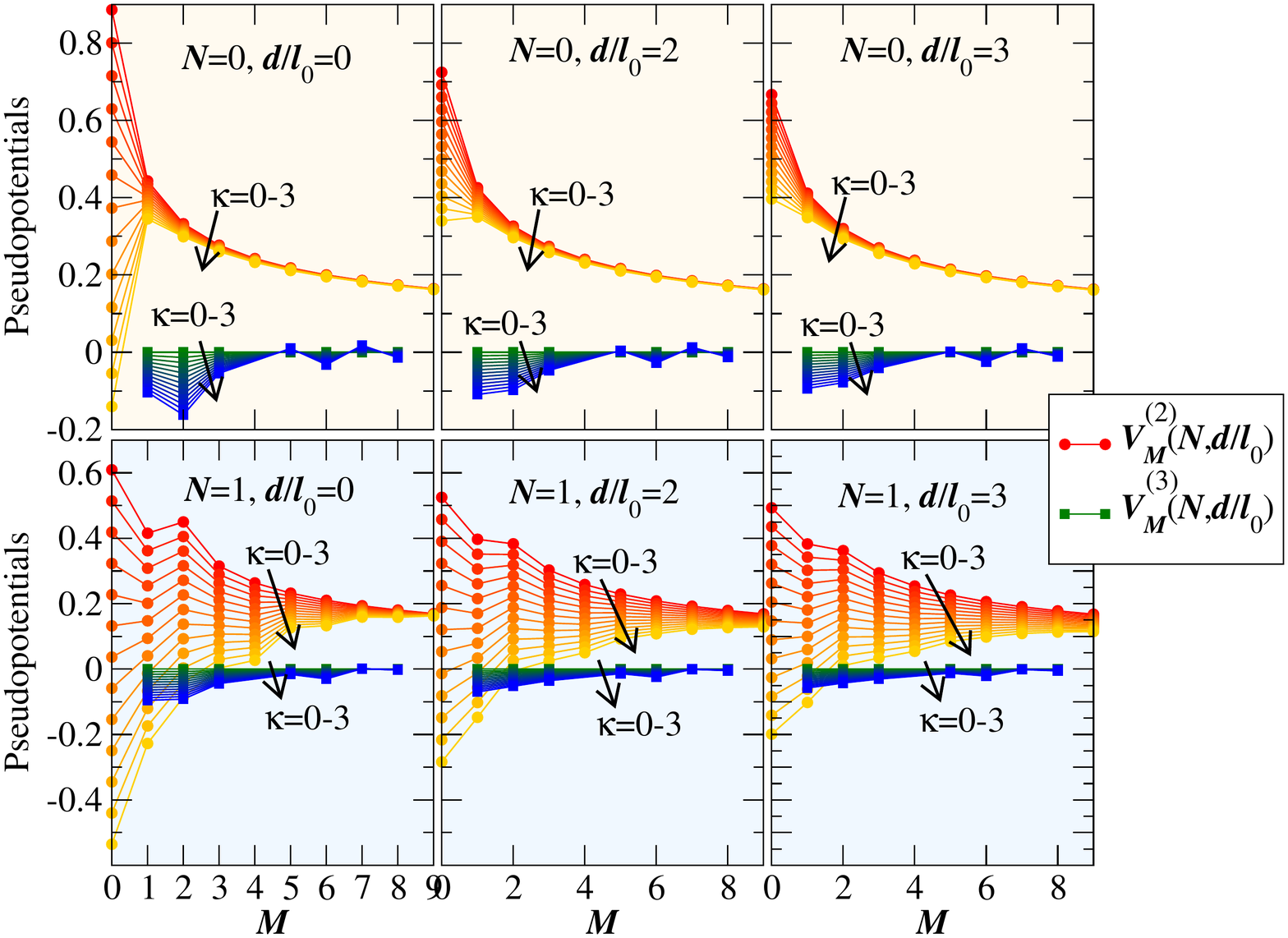}
\caption{(Color online) The two-body and three-body
pseudopotentials, $V^{(2)}_{M}(N,{d/\ell_0})$ and $V^{(3)}_{M}(N,{d/\ell_0})$,
as functions of $M$ for $\kappa=0$-3 for $N=0$ (top panels),
$N=1$ (bottom panels) for $d/\ell_0=0$(left), 2(middle), and 3(right).  Obviously, $V_M^{(2)}(N,d/\ell_0)$ is a sum of the 
bare two-body term plus $\kappa$ times the two two-body corrections.   The two-body pseudopotentials are color-coded such that 
$\kappa=0$ corresponds to red and $\kappa=3$ corresponds to yellow, i.e., the color-coding is ``hot".  The three-body terms 
are color-coded ``cold", i.e., $\kappa=0$ is green and the color changes continuously to blue for $\kappa=3$.  Generally, increasing 
$d/\ell_0$ and $\kappa$ reduces the two-body pseudopotentials compared to the bare values ($\kappa=0$).  The three-body pseudopotentials are 
directly proportional to $\kappa$ so increase in magnitude for increasing $\kappa$, however, increasing $d/\ell_0$ mitigates this 
effect by decreasing the coefficients themselves, as expected.  Please note that any two-body corrections for $M\geq 10$ 
are expected to be small and unimportant.  From inspection of this figure one would expect that Landau level mixing is likely 
to be unimportant for the FQHE in the $N=0$ Landau level for spin-polarized electrons.  For the $N=1$ Landau level, however, it is not 
clear what effect these changes to the pseudopotentials will have on the physics and the answer will have to await future exact diagonalization 
studies.  Of course, even in the $N=0$ Landau level the effect of the three-body terms cold produce non-trivial effects.  All energies 
are given in units of $e^2/\epsilon\ell_0$.}
\label{plot_vms2}
\end{center}
\end{figure*}

The two-body pseudopotential corrections and three-body pseudopotentials
are given in, respectively, Tables \ref{table:2-body-pseudo}
and \ref{table:3-body-pseudo} for $d/\ell_0 = 0$, 1, 2, 3, and 4 and shown 
graphically in Fig.~\ref{plot_vms} up to $d/\ell_0=8$.
The most salient feature is that these corrections are small compared to the bare pseudopotentials
in the $N=0$ Landau level (except for $M=0$) and the $N=1$ Landau level (except for 
$M\leq 3$).  This was already noted in Ref. \onlinecite{Bishara09a} for $d/\ell_0=0$, however, without taking into account 
the modified two-body contribution.
(Note that the numerical values which we find for $d/\ell_0 = 0$ are  different
than in Ref. \onlinecite{Bishara09a} due primarily to the correction to the two-body term coming from 
the correct normal-ordering of the three-body term.)  By ``small compared to the bare 
pseudopotential'', we mean that the coefficients (for $M>0$ in the $N=0$ Landau level and $M\leq 3$ for the $N=1$ Landau level) are $\approx 10$
times smaller and for increasing angular momentum $M$ approximately 50 times 
smaller.   \textit{The smallness of these coefficients means that it
is conceivable that our perturbative calculation is valid even for $\kappa$ beyond one.}
However, in order to say this with certainty, we would need to compute the
order $\kappa^2$, $\kappa^3$, $\ldots$ terms and show that the series converges.

From Fig.~\ref{plot_vms}, we can see the dependence of the two-body corrections 
to the bare pseudopotentials on $d/\ell_0$.  
We  observe in both the lowest and first excited Landau levels that generally the magnitude 
of the corrections to the two-body pseudopotentials decrease with increasing $d/\ell_0$; remember that 
these numbers will be multiplied by $\kappa$ when constructing the full two-body pseudopotentials. 
The bare Coulomb interaction
is itself  also softened as $d/\ell_0$ increases and the fractional change in the bare pseudopotential 
and its two-body correction is similar.   This behavior begins to break down very slightly in the $N=1$ Landau 
level at larger values of $M$ where 
some mild non-monotonic behavior is observed, however, we emphasize that the relative value of the 
correction is quite small for higher $M$'s and this non-monotonic behavior will most likely have 
little effect on the physics.   
The two-body corrections decrease with increasing $d/\ell_0$ and, eventually 
are expected to vanish, so in essence, the effect of finite-thickness is to mitigate 
the effect of Landau level mixing.

In Fig.~\ref{plot_v2_2} we show the two-body corrections to the bare pseudopotentials 
as a function of relative angular momentum $M$ for $d/\ell_0=0$, 1, 2, 3, and 4 for $N=0$ and $N=1$.  As in the case of
the bare pseudopotentials we observe that the two-body corrections decrease with 
both increasing $M$ and $d/\ell_0$.  
However, an effect not seen in the bare pseudopotentials 
for $N=1$ is observed in the two-body corrections; they are larger in magnitude than their $N=0$ counterparts.

Since there is no three-body term at tree-level, the $\mathcal{O}(\kappa)$
correction is the leading three-body term.
Although the three-body interaction is numerically small,
as may be seen in Table \ref{table:3-body-pseudo} and Figs.~\ref{plot_3v_1} and~\ref{plot_3v_2},
it is the leading term which breaks particle-hole symmetry.  The three-body term is especially 
interesting for the FQHE at $\nu=5/2$.  In its absence, particle-hole symmetry is a symmetry
of the Hamiltonian and, consequently, the most promising candidates to describe the 
the 5/2 FQHE, the Moore-Read Pfaffian state
and the anti-Pfaffian state, would be exactly degenerate.  As is well known~\cite{Greiter91,Peterson08c}, 
the Moore-Read Pfaffian is the exact zero-energy solution to a repulsive (positive) three-body Hamiltonian 
that only contains an $M=3$ term.  The anti-Pfaffian, of course, is the exact zero-energy 
solution to the Hamiltonian that is the particle-hole conjugate of the one that yields the Moore-Read Pfaffian.  Under 
particle-hole conjugation, the positive $M=3$ three-body term picks up a minus sign and becomes attractive.   
Additionally, however, a two-body term is also generated under particle-hole conjugation of the 
Moore-Read Pfaffian Hamiltonian.  The 
$M=3$ three-body pseudopotential due to Landau level mixing is negative in both the $N=0$
and $N=1$ Landau levels, a sign which, naively, favors the anti-Pfaffian state over the Moore-Read Pfaffian.
However, with the $M=5$, 6, 7, and 8 three-body terms being non-zero, one cannot predict the physics 
from the pseudopotentials alone.

To illustrate the $d/\ell_0$ dependence of the three-body pseudopotentials,
we show $V^{(3)}_{M}(N,{d/\ell_0})$ as a function of $d/\ell_0$ and $M$ in,
respectively, the $N=0$ and $N=1$ Landau levels in
Figs.~\ref{plot_3v_1} and~\ref{plot_3v_2}. Other than the $M=1$, $N=0$ case
(discussed below), finite width decreases $V^{(3)}_{M}(N,{d/\ell_0})$ 
monotonically.
Figure~\ref{plot_3v_2} shows $V^{(3)}_{M}(N,{d/\ell_0})$ as a function of $M$ and 
we see that for $M\geq 5$ the absolute 
values oscillate: $|V^{(3)}_{5}(N,{d/\ell_0})|<|V^{(3)}_{6}(N,{d/\ell_0})|$,
$|V^{(3)}_{7}(N,{d/\ell_0})|<|V^{(3)}_{6}(N,{d/\ell_0})|$, etc.

We note that there is some non-trivial behavior
in the $M=1$ and 2 three-body pseudopotentials for small values of $d/\ell_0$ for $N=0$, $1$.
The $M=1$ pseudopotential in the lowest Landau level is non-monotonic and starts out
smaller than the $M=2$ pseudopotential.
As $d/\ell_0$ increases, the two cross and $M=1$ is larger than $M=2$ for large $d/\ell_0$.
In the $N=1$ Landau level, we see the same qualitative behavior except that $M=1$ and $2$
do not change signs.

Finally, we examine the overall structure of the effective interaction by showing $V^{(p)}_{M}(N,{d/\ell_0})$
for $N=0,1$ (top and bottom panels, respectively) and
$p=2,3$ in Fig.~\ref{plot_vms2} as a function of $M$ for a range of $\kappa$.  The two- (three-) body pseudopotentials are color-coded 
with ``hot" (``cold") colors going from red to yellow (green to blue) as $\kappa$ is 
changed from 0 to 3.  The  numerical values of the corrections to the bare two-body
pseudopotentials and the three-body pseudopotentials are also small compared to the bare two-body pseudopotentials.
However, as noted in Sec.~\ref{intro},  small quantitative changes to the pseudopotentials
can have large qualitative effects on the possible ground states of the 
effective FQHE Hamiltonian.  As noted above, naively one might expect that the negative value of the $M=3$ three-body term might favor the formation 
of the anti-Pfaffian, but for values of $\kappa>1$ there is a fair amount of renormalization of the two-body pseudopotentials.  It is known from 
previous work~\cite{Peterson08,Peterson08b} that finite thickness stabilizes the Moore-Read Pfaffian at $\nu=5/2$, 
but this calculation 
was only done for $\kappa=0$ and could not distinguish between the Moore-Read Pfaffian and anti-Pfaffian.  From examining the Landau level and sub-band mixing induced  
corrections to the Haldane pseudopotentials, there is potentially a range of $d/\ell_0$ and $\kappa$ that would favor or disfavor the Moore-Read Pfaffian and anti-Pfaffian.  In fact, a cursory look at the two-body pseudopotentials leads one to suspect that Landau level mixing is relatively unimportant for spin-polarized electrons in the $N=0$ 
Landau level, while, evidently, Landau level mixing is much more important in the $N=1$ Landau level.  However, we emphasize that the effect of 
the three-body terms on the physics of the FQHE in either the $N=0$ or 1 Landau levels is non-trivial and not known.  
We
are currently undertaking a thorough numerical study to determine the full quantum phase diagram of this new Hamiltonian for GaAs~\cite{Peterson13b}.  

We now turn our attention to the study of an effective Hamiltonian for graphene.

\section{Effective Hamiltonian for graphene}
\label{sec-graphene}

The formalism for including Landau level mixing in 
graphene (linear dispersion at $B=0$) is similar to that of GaAs,
so we only highlight the differences, such as the absence of finite thickness.
The Landau level 
problem has been investigated by many authors~\cite{Apalkov06,Toke06,Toke07a,Goerbig07,Papic09,Toke07b} 
and we only briefly repeat those results that are relevant to 
the present discussion.

The action [Eq. (1)] for graphene is of the same form as for GaAs with three differences.  
The first, and most important, is that the 
kinetic energy of an electron in the $N^{\rm th}$ Landau level 
is $E^{\rm graph}_{n}=\sgn(n)\sqrt{2|n|}\hbar v_F/\ell_0$ where $v_F$ is the electron Fermi velocity 
($\sim 10^6$ m/s).  Unlike in GaAs, this kinetic energy can 
be positive or negative (electrons or holes) and increases as the square root of the 
absolute value of the Landau level index $n$.  Hence, the 
Landau levels in graphene are not equally spaced; the spacing between successive 
Landau levels decreases (as the inverse square root) as the Landau level index is increased.  The second important difference is that there is no 
sub-band index ``$s$" since the thickness of a graphene monolayer
is atomically thin. The third and final difference lies in the interaction 
matrix elements. We will use the notation $V^{\text{graph}}_{43,21}$ 
for the matrix elements of the Coulomb interaction for electrons in graphene 
and we will continue to use $V_{43,21}$ for the corresponding matrix elements
in GaAs.

\begin{widetext}

In GaAs, the electron-electron interaction matrix element $V_{43,21}$ is given in Eq. (3).  In graphene, Eq. (3) is modified as
\begin{eqnarray}
V^{\text{graph}}_{43,21}&=&\int \frac{d^2k}{2\pi}V(k) e^{-k^2} \frac{\sqrt{2}^{\sum_{i=1}^4\delta_{n_i,0}}}{4}G_{m_4m_2}(-k)
 G_{m_3m_1}(k)\nonumber\\
&&\times \{\sgn(n_4)\sgn(n_2)G_{|n_4|-1,|n_2|-1}(-\bar{k})+G_{|n_4||n_2|}(-\bar{k})\}\nonumber\\
&&\times \{\sgn(n_3)\sgn(n_1)G_{|n_3|-1,|n_1|-1}(\bar{k})+G_{|n_3||n_1|}(\bar{k})\}\;,
\end{eqnarray}
where we set $\sgn(0)=0$ in the above equation. In this equation,
$V(k)={e^2}/(2\pi\epsilon\ell_0 k)$
is the pure (i.e. zero-thickness) Coulomb interaction. 
Note that the definition of $V^{\text{graph}}_{43,21}$ follows from the 
the single-particle energy eigenstates $\eta^{\text{graph}}_{nm}$, which are related to the single-particle eigenstates $\eta_{nm}$ of GaAs according to:
\begin{eqnarray}
|\eta^{\text{graph}}_{nm}\rangle=\left(\frac{1}{\sqrt{2}}\right)^{1-\delta_{|n|,0}}
\left(
\begin{array}{c}
  -\sgn(n) i |\eta_{|n|-1,m}\rangle \\
  |\eta_{|n|m}\rangle  
\end{array}
\right)\;.
\end{eqnarray}

Hence, if $n_i=0$ for $i=1,2,3,4$, then $V^{\text{graph}}_{43,21}=V_{43,21}$.  Thus, 
the bare pseudopotentials for graphene in the $N=0$ lowest Landau level
are identical to those in zero-thickness GaAs.  In the $N^{\rm th}$ Landau level,
the pseudopotenials in graphene are a mixture of the
$N^{\rm th}$ and $(N-1)^{\rm th}$ zero-thickness GaAs Landau levels. 

\end{widetext}

Furthermore, the Landau level index $N$ in graphene is not bounded from below by zero
but instead spans  $-\infty$ to $+\infty$.  Hence, the sum 
in Eq. (11) is modified in the case of graphene to
\begin{eqnarray}
\label{graphsum}
\sum = \sum_{m_x,m_{x'}=0}^\infty \sideset{}{'}\sum_{n_x,n_{x'}=-\infty}^{\infty}\sum_{\gamma,\gamma'=\downarrow,\uparrow}\;.
\end{eqnarray}
Again the prime on the sums over $n_x$ and $n_{x'}$ indicates that we do not include $n_x=N$ or $n_{x'}=N$ in 
the sums.

Another difference between graphene and GaAs is the
 Landau level mixing parameters.  
The Landau level mixing parameter for graphene is $\tilde\kappa=(e^2/\epsilon \ell_0)/(\hbar v_F/\ell_0)=e^2/(\epsilon v_F\hbar)\equiv \alpha_g$ 
where $\alpha_g$ is the graphene fine-structure constant.
Notably, $\tilde\kappa$ is \textit{independent} of magnetic field strength $B$.   
This is in stark contrast to GaAs where $\kappa\propto 1/\sqrt{B}$,
so that Landau level mixing can be ignored for sufficiently large $B$.  
The Landau level mixing parameter in graphene $\tilde\kappa$ is a 
constant that can only be adjusted by manipulating the dielectric constant $\epsilon$, the Fermi velocity $v_F$, or $\hbar$!  In freestanding (i.e., suspended) graphene
$\tilde\kappa\approx 2.2$~\cite{Bolotin09,Du09,Ghahari11}, while for graphene on substrates 
it is $\tilde\kappa\approx 0.9$ for a SiO$_2$ substrate~\cite{Bolotin09} 
and $\tilde\kappa\approx 0.5$--0.8 for a boron nitride substrate~\cite{Dean11,DasSarma-PRB2011}
Thus, by adjusting the dielectric constant of the 
substrate from high dielectric to vacuum, one can only tune the Landau level 
mixing parameter between $0<\tilde\kappa\leq 2.2$ but it cannot be continuously
tuned within a given device by tuning an experimental ``knob" such as the magnetic field strength.  

We also emphasize that since the spacing between Landau levels varies,
unlike in GaAs where it is constant, and we expect the effect of 
Landau level mixing will depend on the Landau level in a manner
that varies approximately as $\propto\tilde\kappa\sqrt{N}$.

\section{THE EFFECTIVE TWO- AND THREE-BODY PSEUDOPOTENTIALS for graphene}
\label{grapheneVms}

We first investigate the three-body pseudopotentials in graphene since the discussion will help facilitate the later 
discussion of the two-body terms.

The most striking feature is that the three-body pseudopotentials vanish for $N=0$.    Recall the expression for the three-body term generated in the perturbative expansion 
of the action for GaAs (Eq.~(\ref{3body})).  For clarity, we write only the sum over Landau level index $n_x$ (and suppress the sums over $m_x$, spin, and cyclic permutations since 
they are not relevant for this argument)
\begin{eqnarray}
&&\sum_{n_x=-\infty,n_x\neq 0}^\infty \frac{\tilde{ V}_{6x,21}^{\alpha'\lambda,\beta\alpha}\tilde{ V}_{54,x3}^{\beta'\gamma',\lambda\gamma}}{\tilde{{E}}^{\rm graph}_{n_x}}\nonumber\\
&&=\frac{\ell_0}{\hbar v_F}\sum_{n_x=-\infty,n_x\neq 0}^\infty \frac{\tilde{ V}_{6x,21}^{\alpha'\lambda,\beta\alpha}\tilde{ V}_{54,x3}^{\beta'\gamma',\lambda\gamma}}{\sgn(n_x)\sqrt{2|n_x|}}\nonumber\\
&&=\frac{\ell_0}{\hbar v_F}\sum_{n_x=-\infty}^1 \frac{\tilde{ V}_{6x,21}^{\alpha'\lambda,\beta\alpha}\tilde{ V}_{54,x3}^{\beta'\gamma',\lambda\gamma}}{-\sqrt{2|n_x|}}
+ \sum_{n_x=1}^\infty \frac{\tilde{ V}_{6x,21}^{\alpha'\lambda,\beta\alpha}\tilde{ V}_{54,x3}^{\beta'\gamma',\lambda\gamma}}{\sqrt{2|n_x|}}\nonumber\\
&&=\frac{\ell_0}{\hbar v_F}\sum_{n_x=1}^\infty\left( \frac{\tilde{ V}_{6x,21}^{\alpha'\lambda,\beta\alpha}\tilde{ V}_{54,x3}^{\beta'\gamma',\lambda\gamma}}{\sqrt{2|n_x|}}
-  \frac{\tilde{ V}_{6x,21}^{\alpha'\lambda,\beta\alpha}\tilde{ V}_{54,x3}^{\beta'\gamma',\lambda\gamma}}{\sqrt{2|n_x|}} \right) \nonumber\\
 &&=0\;\nonumber
\end{eqnarray}
where $\tilde{{V}}_{43,21}^{\beta'\alpha',\beta\alpha}={V}^{\rm graph}_{43,21}\delta^{\alpha\alpha'}\delta^{\beta\beta'}-{V}^{\rm graph}_{34,21}\delta^{\alpha\beta'}\delta^{\beta\alpha'}$ 
and $\tilde{{E}}^{\rm graph}_{n_x}={E}^{\rm graph}_{n_x}-\mu=(\hbar v_F/\ell_0)(\sgn(n_x)\sqrt{2|n_x|}-\sgn(N)\sqrt{2|N|})$.  The canceling of the two terms in the second to last line above can be seen by going back to the definition of ${V}^{\rm graph}_{43,21}$.  Thus, 
the three-body terms exactly vanish in the lowest Landau level ($N=0$).
In the $N=0$ Landau level in graphene, particle-hole symmetry is an exact
symmetry (in the absence of disorder) because there are as many Landau
levels below $N=0$ as above. Consequently, Landau-level mixing \textit{cannot}
generate a three-body term because it would violate particle-hole symmetry.  

\begin{figure}[]
\begin{center}
\includegraphics[width=7.5cm,angle=0]{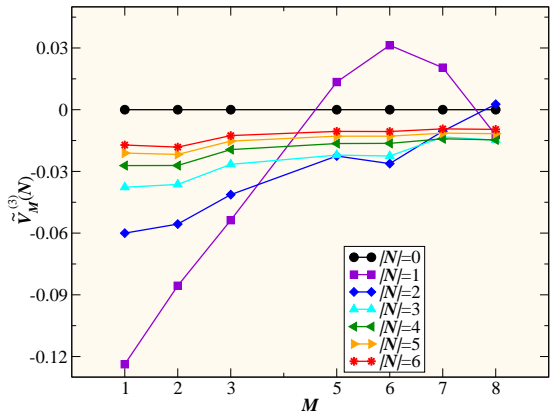}
\caption{(Color online)$\tilde{V}_M^{(3)}(N)$ as a function of $M$ for $N=0$-6.  Note that for $N=0$ all three-body 
pseudopotentials exactly vanish.  All energies 
are given in units of $e^2/\epsilon\ell_0$.}
\label{plot_vms3_graphene}
\end{center}
\end{figure}

Outside of the lowest Landau level ($N\neq 0$), the
three-body pseudopotentials are non-zero. 
They are equal in magnitude and opposite in sign in the $N^{\text{th}}$ Landau level
compared to the $-N^{\text{th}}$ Landau level (see Fig.~\ref{plot_vms3_graphene}).
This, again, follows from particle-hole symmetry.
Furthermore, they are, in general, negative for $N>0$ and, therefore,
positive for $N<0$.

In fact, the Hamiltonian for graphene is invariant under the combined
action of sublattice symmetry, time-reversal, and charge conjugation.
The sublattice symmetry takes $c \rightarrow -c$ on one sublattice,
thereby inverting the kinetic energy in the absence of a magnetic field.
Charge-conjugation transforms particles into holes, thereby inverting
the kinetic energy in the absence of a magnetic field but leaving the
Coulomb interaction energy unchanged (up to a shift of the chemical potential).
Time-reversal inverts the direction of the magnetic field. Therefore,
the combination of all three transformations leaves both the kinetic and
Coulomb energies unchanged. The combination transforms
electrons in the $N^{\text{th}}$ Landau level into holes in the
$-N^{\text{th}}$ Landau level. Therefore, we only give $N\geq0$ results.
The $N<0$ values may be obtained by symmetry, as described above.

Quantitatively, the three-body terms in graphene are similar 
in magnitude to those in GaAs, small compared to the
bare two-body pseudopotentials, and decreasing
in magnitude with increasing $M$.
For $|N|=1$ there is non-trivial, non-monotonic $M$-dependence,
where the pseudopotentials for $M=5$, 6, and 7 are opposite in sign to
those for $M=1$, 2, 3, and 8.  Table~\ref{table:3-body-pseudo-graphene} lists
 the numerical values of $\tilde{V}^{(3)}_M(N)$ for $|N|=0-4$ and $M=1-3$, and $5-8$.

\begin{table}[t]
\caption{We list the numerical values of the three-body pseudopotentials $\tilde{V}^{(3)}_{M}(N)$
as a function of $M$ for graphene.  All energies 
are given in units of $e^2/\epsilon\ell_0$.
}
\begin{ruledtabular}
\label{table:3-body-pseudo-graphene}
\begin{tabular}{lrrrrr}
$N=$  & 0 &  1  & 2 & 3 & 4 \\ \hline
$ \tilde V_{1}^{(3)}(N)$ & 0 &  -0.1237  & -0.0600  & -0.0377 & -0.0272 \\
$ \tilde V_{2}^{(3)}(N)$ & 0 &  -0.0856  &  -0.0556 &  -0.0363 & -0.0271 \\
$ \tilde V_{3}^{(3)}(N)$ & 0 &  -0.0537  & -0.0413  & -0.0266 & -0.0194 \\
$ \tilde V_{5}^{(3)}(N)$ &  0 &  0.0135  & -0.0225 & -0.0221  & -0.0165 \\
$ \tilde V_{6}^{(3)}(N)$ & 0 &   0.0313  & -0.0262 & -0.0225  & -0.0164 \\
$ \tilde V_{7}^{(3)}(N)$ &  0 &  0.0205  & -0.0105 & -0.0133  & -0.0141 \\
$ \tilde V_{8}^{(3)}(N)$ & 0 &  -0.0123  &  0.0027 & -0.0148  & -0.0147 \\
\end{tabular}
\end{ruledtabular}
\end{table}

\begin{table}[t]
\caption{We list the numerical values of $\delta \tilde{V}^{(2)}_{M}(N)$ for $N=0$, 1, and 2.
The two-body correction consists of  two contributions:  one is a direction calculation of the ZS, ZS', and BCS 
diagrams and the other arises from the normal ordering of the three-body term [Eq.~(\ref{eqn:2bodyfinal}], please see the discussion in the text above.  
In the $N=0$ Landau level, these corrections are similar in magnitude to the results for GaAs.  However, for 
$N=1$ and 2, the corrections are much larger and, in fact, it is expected that effective Landau level mixing parameter 
$\tilde\kappa$ is really $\propto\tilde\kappa\sqrt{N}$, see discussion in text.  
 We use the $N=0$ values as the first term contribution 
 for all $N$ since this term is so much smaller in magnitude than the second term and expected to make little difference to the final values or the physics.  All energies 
are given in units of $e^2/\epsilon\ell_0$.}
\begin{ruledtabular}
\label{table:2-body-pseudo-graphene}
\begin{tabular}{lrrr}
 & $N=0 $ &  1  & 2  \\ \hline  
$\delta \tilde{V}_{0}^{(2)}(N)$ & -0.2638 & -0.4519 & -0.8529    \\ 
$\delta \tilde{V}_{1}^{(2)}(N)$ & -0.0633 & -0.1762 &  -0.5240    \\
$\delta \tilde{V}_{2}^{(2)}(N)$ & -0.0407 & -0.0690 &  -0.3943   \\ 
$\delta \tilde{V}_{3}^{(2)}(N)$ & -0.0143 & -0.0290 &  -0.3010  \\ 
$\delta \tilde{V}_{4}^{(2)}(N)$ &  -0.0090 & -0.0123&  -0.2179  \\ 
$\delta \tilde{V}_{5}^{(2)}(N)$ &  -0.0052 & -0.0066  & -0.1510 \\
$\delta \tilde{V}_{6}^{(2)}(N)$ & -0.0034 & -0.0032 &  -0.0996   \\
$\delta \tilde{V}_{7}^{(2)}(N)$ & -0.0030 & -0.0022 & -0.0643    \\
$\delta \tilde{V}_{8}^{(2)}(N)$ & -0.0022 & -0.0012 &  -0.0400   \\
$\delta \tilde{V}_{9}^{(2)}(N)$ & -0.0016 & -0.0005 &   -0.0242  \\
\end{tabular}
\end{ruledtabular}
\end{table}

\begin{figure}[]
\begin{center}
\includegraphics[width=7.5cm,angle=0]{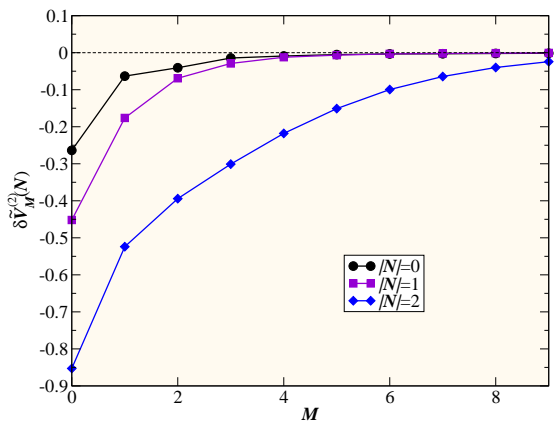}
\caption{(Color online) The two-body pseudopotential corrections, $\delta \tilde{V}_M^{(2)}(N)$ for $N=0$, 1, and 2 for 
graphene as a function of $M$.  It is clear from this figure that the pseudopotential corrections increase for increasing Landau 
level $N$ reflecting the conjecture that the effective Landau level mixing parameter is actually $\propto\tilde\kappa\sqrt{N}$ and 
not simply $\tilde\kappa$.  All energies 
are given in units of $e^2/\epsilon\ell_0$. }
\label{plot_v2_graphene}
\end{center}
\end{figure}

\begin{figure*}[htbp]
\begin{center}
\includegraphics[width=14.cm,angle=0]{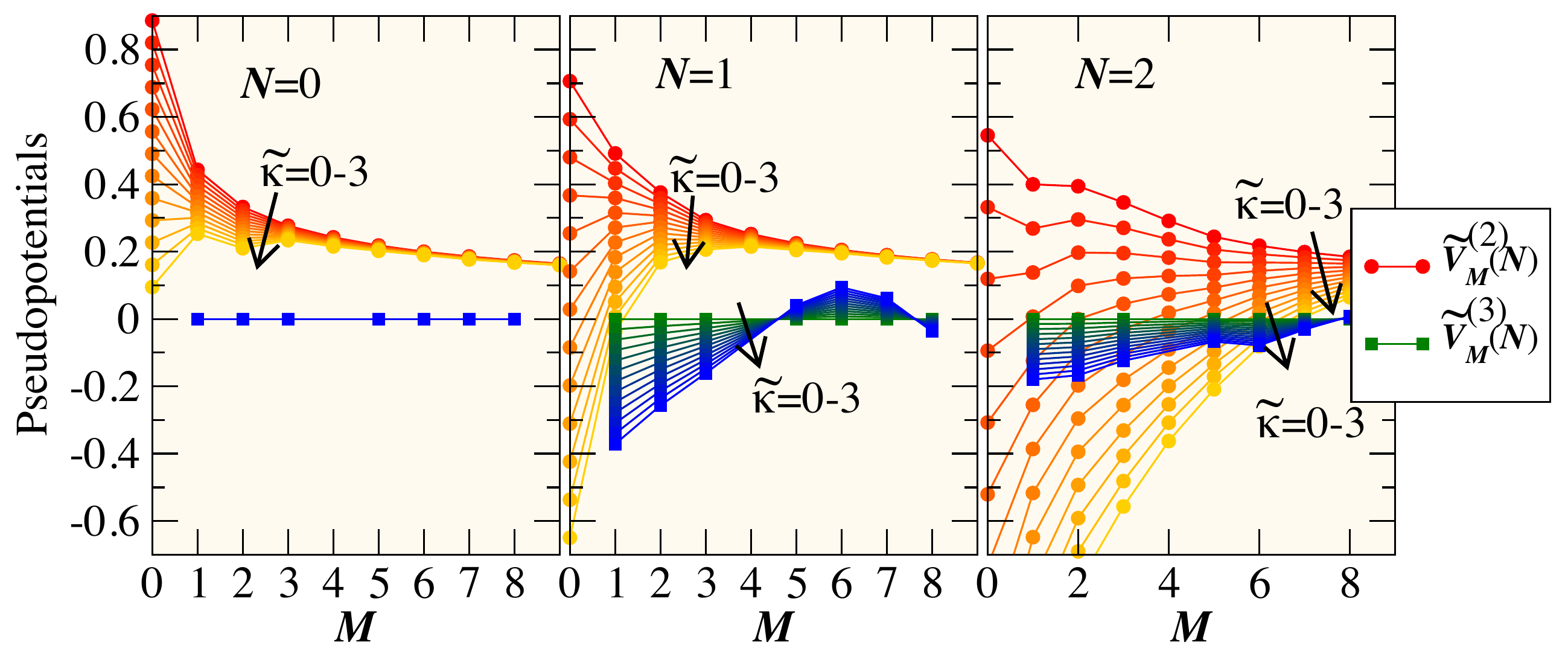}
\caption{(Color online) Both the two- and three-body
pseudopotentials as functions of $M$ for $\tilde\kappa=0-3$ for graphene, in the $N=0$, 1, and 2 Landau levels.  For the lowest Landau level ($N=0$) 
there is no three-body term (it exactly vanishes) and the two-body corrections are modest.  Hence, one would expect, from merely studying the 
pseudopotentials, that the FQHE in $N=0$ Landau level for graphene would be nearly identical to that of GaAs.  However, for $N\neq0$, we 
observe significant corrections to the two-body pseudopotentials and significant three-body terms.  It is not at all clear what these effects 
will have on the FQHE.  Naively, one is tempted to suggest that the FQHE will not be observed in $N\neq 0$ Landau levels in graphene, or if it is 
observed, it will be of an exotic variety.  All energies 
are given in units of $e^2/\epsilon\ell_0$.}
\label{plot_vms2_graphene}
\end{center}
\end{figure*}

Turning now to the two-body pseudopotentials, we note first that because the energy increases
with Landau level index $n$ as $\sqrt{n}$, the contribution from higher
levels will be less strongly suppressed. In fact, unlike in the case of GaAs,
it is extremely helpful to use the constraint following from angular momentum
conservation to eliminate one of the Landau level index summations
in Eqs.~(\ref{eqn:ZS})-(\ref{eqn:BCS}), thereby making the convergence
of the sum more apparent.  

The contribution to the two-body term coming 
from the normal ordering is usually an order of magnitude larger than the term 
coming from the ZS, ZS', and BCS diagrams alone.  Because of this fact, and 
the fact that finding the convergence of the smaller two-body terms is  laborious, we use the $N=0$ 
values of the two-body terms for all Landau levels as an upper limit and good approximation.  Table~\ref{table:2-body-pseudo-graphene} lists
the numerical values of $V^{(2)}_M(N)$.

In addition, in Fig.~\ref{plot_v2_graphene} we plot the Landau level mixing induced corrections to the 
two-body pseudopotentials as a function of relative angular momentum $M$ for $N=0$, 1, and 2.  
The two-body corrections are universally negative and for small $M$ can be 
approximately half the value of  the bare pseudopotential.  Interestingly, the corrections become larger in higher Landau levels 
owing to the fact that the effective Landau level mixing parameter is $\propto\tilde\kappa\sqrt{N}$.  

In Fig.~\ref{plot_vms2_graphene} we show the two- and three-body pseudopotentials as a function 
of both $\tilde\kappa$ and $M$.   For $N=0$,  Landau level mixing has little 
effect on the pseudopotentials, i.e., the corrections to the two-body term are relatively small and there is no three-body term--this might help 
explain why the FQHE 
is observed in the $N=0$ Landau level in graphene, when the system is spin and valley polarized, even though Landau level mixing cannot 
naively be ignored due to a large value of $\tilde\kappa$.  
This further underscores the fact that the FQHE 
in graphene in the $N=0$ is observed to be, and is expected to be, nearly identical to that of GaAs (other than 
interesting complications that arise when the spin and valley degeneracies remain).  However, for $N\neq 0$, there are strong corrections 
in the form of negative two-body corrections and emergent three-body terms.  For moderate values of $\tilde\kappa$ 
it is not at all clear that graphene would even exhibit the FQHE according to our calculations--the two-body terms can be of the same order of magnitude 
as the three-body terms.  We also emphasize that with such strong Landau level mixing corrections to first order in $\tilde\kappa$ it 
is possible that higher order corrections will be significant, however, it is likely that these corrections (which we are ignoring) will 
make the situation worse for the FQHE, not better.  

Of course, it remains to be seen theoretically what effect the three-body 
terms have on the FQHE.  Exact diagonalization studies using the
above effective Hamiltonian are likely to shed light on this
problem~\cite{Peterson13b}.

\section{Conclusion}
\label{conclusion}

Quantum electrodynamics is a paradigm of a perturbative
theory: the dimensionless expansion parameter $\alpha\approx 1/137$
is small, so effects that occur at each order in perturbation theory
are successively less important than those that occurred at the previous
order. However, in condensed matter physics, we have grown accustomed to
perturbation theory in dimensionless parameters that are not particularly
small. Landau parameters are typically $\mathcal{O}(1)$, yet Fermi liquid theory
remains valid for $^3$He and, often, for electrons in metals. The ratio of the
Coulomb energy to the kinetic energy in a metal, $r_s$, can
be moderately large, without the metallic state being destabilized
(in favor of, say, a Wigner crystal). But we should guard against the
possibility that we have been lulled into a false sense of security.
There is {\it no} quantum Hall experiment in which Landau-level mixing
is obviously negligible {\it a priori}. In the original observation of the
fractional quantum Hall effect at $\nu=1/3$, the Landau level mixing
parameter is $\kappa=0.65$. This might be small enough that one
can reach this value via an expansion
in powers of $\kappa$ about $\kappa=0$ or it might not,
but it is certainly not a very small number such as $1/137$
which one {\it a priori} expects to be negligible.
In the $N=1$ Landau level, $1<\kappa<2.5$, so it is even less clear
that Landau level mixing is small.
In the case of graphene, $0.5<\tilde\kappa<2.2$, so the situation is no better.

The coefficient of $O(\kappa)$ corrections to two- and three-body pseudopotentials
due to Landau level mixing are almost always
numerically small.
Consequently, these corrections are small even at $\kappa=1$.
(Most of the corrections are small even at $\kappa=10$.)
Since we have not computed the order $\kappa^2$ corrections,
this does not prove that experiments are effectively in a small $\kappa$
regime. However, it does, at least, raise this possibility -- or, in other words,
the possibility that the true expansion parameter is $\sim\kappa/10$.

Of course, for the FQHE in the $N\neq 0$ Landau level of graphene, the corrections
are not necessarily small and it is very possible that a 
perturbative treatment of Landau level mixing is inappropriate.  

Strictly speaking, our computation yields an effective action,
rather than an effective Hamiltonian -- as is always the case
when integrating out high-energy states to produce a low-energy
effective theory. This effective action has retarded interactions but, at lowest-order
in $\kappa$, these retarded interactions can be neglected. Moreover,
our calculation assumed fixed chemical potential. At lowest-order
in $\kappa$, we can simply take this to be the Landau level energy
$\hbar \omega_c (N+\frac{1}{2})$ or $\sgn(N)\sqrt{2|N|}\hbar v_F/\ell_0$,
regardless of filling fraction within the $N^{\text{th}}$ Landau level.
However, in going to higher-order in $\kappa$, we would have to
tune the chemical potential to determine the filling fraction.
For these two reasons, it will be difficult to determine the ground state
of the system to order $\kappa^2$. But it should still be possible
to derive an effective action (albeit with retarded interactions)
from which it would be possible to determine if order $\kappa^2$
effects are also small even at $\kappa=1$.

We note that, for large numbers of electrons, fixing the
chemical potential will be the same as fixing the electron
number (or, equivalently, the filling fraction). However, for the small numbers
of electrons considered in numerical exact diagonalization studies, fixed chemical
potential and fixed electron number are not the same.
For fixed chemical potential, $\Delta n \propto \sqrt{n}$ (where
$n$ is the electron number) which can be large. Thus, our effective
Hamiltonian is only, strictly speaking, correct for systems with a large number of
electrons, not a small number. However, if the goal of studying
small systems is to extrapolate
their properties to larger ones to determine the solution in the
thermodynamic limit with fixed filling fraction,
then our effective Hamiltonian can be used.

Even if it is quantitatively small, Landau-level mixing could,
nevertheless, have significant qualitative effects. It is the leading effect
that breaks particle-hole symmetry within a Landau level (except
for the $N=0$ Landau level in graphene, where this is an exact symmetry).
This has important effects at $\nu=5/2$ in GaAs, where it could break the degeneracy between
the Moore-Read Pfaffian state and the anti-Pfaffian state, as we discuss elsewhere~\cite{Peterson13b}.
It can also have important effects at $\nu=7/3, 8/3$ and $12/5, 13/5$--with particle symmetry broken, these states may not be
particle-hole conjugates of each other after all but, instead, have very different
characters.
Indeed, even within the lowest Landau level, there are obvious differences
between particle-hole conjugate plateaus, such as $\nu=1/5$ and $\nu=4/5$.
Presumably the underlying cause is Landau level mixing.

Finally, we note that, although the corrections to the effective
Hamiltonian due to Landau level mixing are small compared to the
bare terms, they are not small compared to the energy gaps
of quantum Hall states. So it is possible, in principle, for three-body terms to have a non-negligible effect on the energy
gaps of some fractional quantum Hall states while, at the same time,
order $\kappa^2$ terms can be neglected. In such a case, our order $\kappa$
effective Hamiltonian would be the most promising starting point
for an attempt to make a direct quantitative comparison between
experiments and numerical simulation of quantum Hall systems.

Note: Recently, we received
a first draft of a paper by Sodemann and MacDonald~\cite{Sodemann13} 
containing some results on the effect of Landau level mixing
on the FQHE in GaAs in the zero-thickness limit. They reproduced
the $d/\ell_0=0$ limit of our three-body interaction (which, in turn, is equal
to the three-body interaction of Bishara and Nayak \cite{Bishara09a})
and reproduced the $d/\ell_0=0$ limit of our two-body interaction
in the $N=0$ Landau level. However our two-body interaction in the
$N=1$ Landau level diverges from theirs. In contrast, the $d/\ell_0=0$ limit of
our two- and three-body interactions in both the $N=0$ and $N=1$ Landau levels
are within a few percent of the results of Rezayi and Simon \cite{Rezayi13}.

\begin{acknowledgements}
We acknowledge DARPA QuEST and Microsoft for funding and M. R. P. acknowledges 
California State University Long Beach Start-up Funds.  We acknowledge
helpful discussions with Matthias Troyer and Matt Hastings and, additionally, Euyheon Hwang, Shaffique Adam, and Sankar Das Sarma regarding the effective dielectric constant for graphene 
on various substrates.
We benefited from stimulating and helpful discussions with
Steve Simon, Ed Rezayi, and Inti Sodemann. We are especially
grateful to  Simon and  Rezayi for sharing their, at the time, unpublished results \cite{Rezayi13}.  
We thank the Aspen Center for Physics, where part of this
work was completed, for its hospitality.  Last, we thank one of the referees for helpful comments that improved the manuscript.
\end{acknowledgements}


\begin{thebibliography}{86}
\bibitem[{foo({\natexlab{a}})}]{footnote1}
\bibinfo{note}{One may also question whether the same ground state is observed
  in all experiments, given that the $\nu=5/2$ state is observed from $B=2$T
  all the way up to $B=12$T (which corresponds to the range $0.7 < \kappa <
  1.8$) and is observed in both quantum wells and heterostructures -- implying
  that the Hamiltonian varies considerably among the various experiments which
  have observed a $\nu=5/2$ plateau.}
\bibitem[{foo({\natexlab{b}})}]{footnote2}
\bibinfo{note}{Most of these numerical studies also assume that the ground
  state is fully spin-polarized. The few which do not make such an assumption,
  in fact, make a strong case for a spin-polarized ground state
  \cite{Morf98,Dimov08,Feiguin09}. However, experiments indicate that the
  ground state may not be polarized \cite{Stern10,Rhone10}.}
\bibitem[{foo({\natexlab{c}})}]{footnote3}
\bibinfo{note}{Interferometry experiments~\cite{Radu08,Dolev08,Willett09} can
  distinquish between a $\nu=5/2$ state described by the MR Pfaffian or
  anti-Pfaffian.}
\bibitem[{foo({\natexlab{d}})}]{footnote-sodemann}
\bibinfo{note}{The corrections for $d/\ell_0=0$ were recently also reported in
  Ref.~\onlinecite{Sodemann13} for the $N=0$ and 1 Landau levels. Their numbers
  agree with ours in the $N=0$ Landau level but not the $N=1$.}
\end{thebibliography}

\end{document}